\documentclass[aps,prc,preprint,showpacs,showkeys,floatfix,nofootinbib,%
superscriptaddress]{revtex4}
\usepackage{graphicx}
\usepackage{multirow}
\usepackage{bm}

\def\be{\begin{equation}} \def\ee{\end{equation}} \def\bea{\begin{eqnarray}}
\def\eea{\end{eqnarray}} \def\nnb{\nonumber}

\newcommand{\eqn}[1]{\label{eq:#1}}
\newcommand{\refeq}[1]{(\ref{eq:#1})}

\newcommand{\Eq}{Eq.~\refeq}
\newcommand{\Eqs}[2]{Eqs.~(\ref{eq:#1}) and (\ref{eq:#2})}

\newcommand{\vecbox}[1]{\mbox{\boldmath{$#1$}}}

\def\1s0{{}^1S_0}
\def\ts1{{}^3S_1}
\def\td1{{}^3D_1}
\def\ltap{\ \raise.3ex\hbox{$<$\kern-.75em\lower1ex\hbox{$\sim$}}\ }
\def\gtap{\ \raise.3ex\hbox{$>$\kern-.75em\lower1ex\hbox{$\sim$}}\ }
\def\ket#1{\vert#1\rangle}

\def\braketa#1{\langle#1\rangle}

\newcommand{\EG}{{\textit{e.g.}}}
\newcommand{\IE}{{\textit{i.e.}}}

\date{\today}

\begin{document}

\title{Consistency between renormalization group running of chiral
operator and counting rule\\
---  Case of chiral pion production operator  ---}

\author{Satoshi X. Nakamura}\email{satoshi@jlab.org}
\thanks{Current affiliation: 
Excited Baryon Analysis Center (EBAC),
Thomas Jefferson National Accelerator Facility, Newport News, VA 23606, USA}
\affiliation{Theory Group, TRIUMF,
4004 Wesbrook Mall, Vancouver, BC V6T 2A3, Canada}
\affiliation{Instituto de F{\'i}sica, Universidade de S{\~a}o Paulo,
S{\~a}o Paulo -SP, 05508-090, Brazil}

\begin{abstract}
In nuclear chiral perturbation theory ($\chi$PT), an operator is defined
in a space with a
cutoff which may be varied within a certain range.
The operator runs as a result of the variation of the cutoff
[renormalization group (RG) running].
In order for $\chi$PT to be useful,
 the operator should run
in a way consistent with the counting rule;
that is,
the running of chiral counter terms have to be of natural size.
We vary the cutoff using
the Wilsonian renormalization group (WRG) equation,
and examine this consistency.
As an example, we study the $s$-wave pion production 
operator for $NN\to d\pi$, derived in $\chi$PT.
We demonstrate that the WRG running does not generate any
chiral-symmetry-violating (CSV) interaction, provided that we
start with an operator which does not contain a CSV term.
We analytically show how the counter terms are generated in the WRG
running in case of the infinitesimal cutoff reduction.
Based on the analytic result, we argue
a range of the cutoff variation for which 
the running of the counter terms is of natural size.
Then, we numerically confirm this.
\end{abstract}


\maketitle

\section{Introduction}\label{sec_intro}

In nuclear chiral perturbation theory ($\chi$PT), operators (\EG,
nuclear force, electroweak current and pion production operator)
are derived from a chiral Lagrangian following a counting rule.
Many processes in few-nucleon system have been successfully described by
these chiral operators, establishing the validity of this approach.
\cite{review_eft,kubodera_review}
The chiral nuclear operators
are defined in a model space with a certain cutoff.
The cutoff has a physical meaning and its choice is not
arbitrary.
Given a long-range (\EG, pion exchange) mechanism, 
the cutoff should be smaller than the scale where the details of
shorter-range mechanisms are resolved.
For example,
in describing a nuclear operator with pion-exchange mechanisms plus
contact terms,
the cutoff should be smaller than the $\rho$ meson mass.
Also, the cutoff should not be so small that the long-range mechanism is
not fully taken into account.
Although not arbitrary,
the cutoff still may be varied within a physically reasonable region.
As a result of the variation of the cutoff, an operator 
(more specifically, couplings of counter terms)
runs so that observables are cutoff-independent
[renormalization group (RG) running].
A question here is whether the running of the counter terms is
consistent with the counting rule.
Stated differently, we wonder whether the counter terms run with
keeping the size of the couplings natural [${\cal O}(1)$].
This consistency is a necessary condition for the counting rule to be
useful because, if not satisfied, the counting rule does not correctly 
reflect the ordering of the importance of the counter terms.

In order to address the question
concerning the consistency between the RG running and the counting rule,
we propose to use a RG equation which controls the running of an
operator so that Green's functions are unchanged.
Wilsonian RG (WRG) equation is such an equation.
For studying the RG running, we consider it the most
appropriate to use the WRG equation because
we have the following beneficial points:
(i) The WRG equation is consistent with the derivation of
the effective Lagrangian with a path integral, see Sec.~\ref{sec_wrg};
(ii) It is guaranteed that no chiral-symmetry-violating (CSV)
interactions are generated by the WRG running, 
\footnote{
In Sec.~\ref{sec_wrg}, 
we will explain in more detail what we mean by ``no CSV terms
are generated in the WRG running''.
}
provided that no CSV term exists
(\EG, the operator vanishes at threshold in the chiral limit)
in both a transition operator and a nuclear force
before the RG running, see Sec.~\ref{sec_counter};
(iii) Since we have
the RG running of the operator which correctly reflects the
high-momentum states integrated out,
we can study
the convergence of the chiral expansion of the RG running.
Our usage of the WRG equation is different from
Refs.~\cite{birse,HaradaKubo} in which
the scaling behavior of each interaction
near the fixed point is identified with power counting.
What we would like to do is also different from Refs.~\cite{nogga,epel}
where the renormalizability of the leading order (LO) chiral $NN$ potential 
over a very wide range of the cutoff is examined.
We apply the WRG equation to a set of operators derived with an established
power counting, and then examine the internal consistency between 
the RG running and the counting rule
over a physically reasonable range of the cutoff.
In the above RG analyses\cite{birse,HaradaKubo,nogga,epel},
the authors are {\it not} primarily interested in the actual size of the
running for a physically reasonable range of the cutoff.
We are interested in the actual size of the RG running and its
consistency with the counting rule.
One may naively expect that this kind of calculation only generates the
counter terms which scale as $\propto 1/\Lambda$ ($\Lambda$ : cutoff).
We will semi-analytically show that this expectation is not always the case,
and find a size of the variation of $\Lambda$ for which this expectation
holds true.
Also, we argue that
the RG running is consistent with the counting rule 
if we vary the cutoff smaller than this size;
otherwise, the RG running could be divergent.
This argument is also consistent with
the divergent RG running of some contact $NN$ interactions 
which was found in Ref.~\cite{nogga} 
(\EG, see Fig.~2 of the paper).
The physically reasonable range of the cutoff is consistent with 
the size of the variation of $\Lambda$ which we semi-analytically
find.

We quantitatively confirm the above qualitative analysis by a numerical
calculation.
\footnote{Similar analyses addressing the RG running and the
naturalness of the counter terms in the $NN$ interaction are found in 
Refs.~\cite{rg1,epel2};
the WRG equation is used in Ref.~\cite{rg1} while a unitary
transformation is used in Ref.~\cite{epel2}.}
In the numerical calculation, we use a phenomenological nuclear force
rather than a $\chi$PT-based nuclear force.
A reason is that
chiral $NN$-potentials available\cite{chiral_potential} have relatively
small cutoffs ($\Lambda \sim$ 500~MeV) and are
not very appropriate to study the WRG running in the cutoff range
considered in this work (500~MeV $ \le \Lambda \le$ 800~MeV).
Phenomenological $NN$-potentials such as the CD-Bonn
$NN$-potential~\cite{cdbonn} 
which we use are not based on the chiral Lagrangian.
However, these potentials have been often used with chiral operators
(electroweak currents, pion production operators) to calculate matrix
elements (hybrid approach).
This hybrid approach has been used extensively, through which
its usefulness has been established\cite{kubodera_review}.
Despite the phenomenological success of the hybrid approach,
there is still a concern about the consistency between
the nuclear force and the chiral transition operators.
In this context,
the present RG analysis may provide an interesting test for the hybrid
approach.
%
%
This is because
if a CSV term is generated in the WRG running,
then the result signals that the nuclear force contains a numerically
significant CSV component and
one has to doubt the validity of the hybrid approach.

This paper is organized as follows.
In Sec.~\ref{sec_wrg}, 
we first develop a formal apparatus to address the question
of how to evolve the operator for changing $\Lambda$, using the WRG equation.
In Sec.~\ref{sec_op}, 
we present expressions for 
the chiral next-to-leading order
(NLO) $s$-wave $\pi$-production operator~\cite{lensky}
to which 
we apply the WRG analysis
as a demonstration, and address the question of the
consistency.
The operator we use is
based on the heavy-baryon $\chi$PT with a revised power counting 
scheme~\cite{mod_count} which treats the large incoming nucleon momentum as 
a separate large energy/momentum scale (see, \EG, Ref.~\cite{DKMS} for a 
treatment using the original Weinberg counting~\cite{Weinberg}).
We use the revised power counting throughout this work unless otherwise
specifically stated.
In Sec.~\ref{sec_counter},
we reduce the (sharp) cutoff of the chiral $\pi$-production operator
by the infinitesimal amount using the WRG equation and the chiral LO
nuclear force.
We obtain analytic expressions which show that
the WRG equation indeed induces the running of the chiral counter terms;
no CSV terms are generated.
%
In Sec.~\ref{sec_result}, 
in order to quantitatively address the RG running,
we reduce the cutoff of the $\pi$-production operator
from $\Lambda$ = 800~MeV down to $\Lambda$ = 700, 600 and 500~MeV
using the WRG equation with a phenomenological nuclear force.
As a result we obtain numerically 
the corresponding effective operators.
Then, we try to reproduce the effective RG low-momentum
operator\footnote{
In this work, we will use
the term ``low-momentum interaction'' to refer to an effective
interaction obtained with the Wilsonian renormalization group equation.}
 by calculating another operator consisting of the chiral
NLO operator 
and the chiral counter terms; the chiral counter terms are introduced
following the counting rule.
The counter terms are accompanied by unknown coefficients, the
so-called low-energy constants (LECs), which are determined by fitting
to the low-momentum operator.
We will show that the running of the operator with $\Lambda$ is, to a very high
precision, captured by the lowest order counter term.
Finally, we summarize our findings in Sec.~\ref{sec_summary}.
Some calculations are relegated to an Appendix.

\section{Wilsonian renormalization group equation for transition
 operator}
\label{sec_wrg}

The WRG equation is an equation which is designed to control
the running of a set of operators,
as a result of the cutoff reduction,
so that Green's functions are unchanged.
At first, we state why we use the WRG equation to reduce the cutoff.
An effective Lagrangian can be obtained formally via a path integral 
formulation based on the Lagrangian of the underlying, more fundamental theory.
One integrates out the high energy degrees of freedom using the path integral.
When integrating out the high momentum states of the nucleon in the 
heavy-baryon $\chi$PT Lagrangian, we can also use the path integral.
This procedure is equivalent to solving the WRG equation derived below;
we will state more on this later.
Thus one way of reducing the cutoff consistently with 
effective field theory is through using the WRG equation.
It is not a priori guaranteed that the WRG running of
the counter terms is consistent with the counting rule,
and thus we consider it worthwhile examining.
We will see that the WRG running is consistent when
the cutoff is changed within a physically reasonable range;
otherwise, not necessarily consistent.

Contrary to the approach in the above paragraphs, some
previous works did not use a RG equation for studying RG running.
Instead, for a given cutoff, the (leading) counter term was fixed so that
some observables were reproduced.
Repeating this procedure over a certain range of the cutoff leads to
the RG running of the counter term.
In this procedure, it is assumed that the running of the (leading)
counter term simulates the high momentum states integrated out.
However, one could fix the coupling of a higher-order counter term,
instead of the leading term, so that some observables
were reproduced.
Without using a RG equation,
it is impossible to know which counter term is running.
Furthermore, our approach based on the WRG equation is more advantageous
than the above RG analyses on the following two points:
(i) it is guaranteed that no CSV operator is 
generated in the RG running, provided that no CSV term exists
before the RG running; 
(ii) the RG running of the operator which exactly reflects the high momentum
states integrated out is at hand, and thus we can examine
whether the RG running is captured by a series of the chiral counter terms
consistently with the counting rule.
We detail the above statement (i) in the following.
The chiral operators in the nuclear $\chi$PT
are not chiral invariant in the sense that the original Lagrangian is
invariant under chiral transformations. 
We take advantage of the chiral symmetry by deriving the operators from
the chiral Lagrangian; the parameterization of the operators is given by
the chiral Lagrangian.
After the WRG running, if the operators are still accurately
parameterized by the same parameterization, 
we state that no CSV interactions are generated.
We do not address 
the chiral symmetry invariance of the original chiral Lagrangian
in the RG running.
Here we are interested in
nuclear operators based on $\chi$PT,
which is actually used in practical calculations,
and their WRG running.
As far as we examine the WRG running of the $\chi$PT-based operators
with the cutoff, no CSV term is generated.

As stated in the previous paragraph, the WRG equation can be derived using 
the path integral. 
The present author derived the WRG equation for the $NN$
interaction in this way in Ref.~\cite{rg1}.
The WRG equation for a transition operator ($\pi$ production operator in
our case) can also be derived  in essentially the same way, whereas we
can also derive it in a simpler way as detailed in Ref.~\cite{NA}
(Appendix~A of that reference).\footnote{
In Ref.~\cite{NA}, the WRG equation was derived as a sufficient condition
of the cutoff independence of a matrix element.
Recently, it was shown that the same WRG equation is also
necessary-sufficient condition, 
if one imposes the cutoff independence of
the five point Green function~\cite{KB}.}
Here we derive the WRG equation following Ref.~\cite{NA} for simplicity
{\it without} using the partial wave decomposition 
that was used in Ref.~\cite{NA}.
We start with a matrix element in which the transition operator is defined in 
a model space spanned by plane wave states of the two-nucleon system.  
The maximum magnitude of the relative momentum in the model space is
given by the cutoff, $\Lambda$.
The matrix element of an operator $O$ is given by
\begin{eqnarray}
\eqn{o_me}
\braketa{\bm{p}'|O|\bm{p}}
= \int_0^\Lambda\!\! d\bm{k}
\int_0^\Lambda\!\! d\bm{k}'\;
\psi_{\bm{p}'}^\dagger(\bm{k}')\; O(\bm{k}',\bm{k})\; \psi_{\bm{p}}(\bm{k}) \ ,
\end{eqnarray}
where $\psi_{\bm{p}} (\bm{k})$ [$\psi_{\bm{p}'}(\bm{k}')$] is
the wave function for the initial (final) two-nucleon state.
The wave functions are derived from a low-momentum $NN$
interaction with the same cutoff $\Lambda$.
The quantity $\bm{p}$ ($\bm{p}'$) is the on-shell relative momentum for 
the initial (final) two-nucleon state.
The on-shell momentum is related to the energy ($E$) for the relative motion 
of the two nucleons through $p = |\bm{p}| = \sqrt{ME}$, where $M$ is the nucleon mass.
We denote the transition operator $O(\bm{k}',\bm{k})$, where
$\bm{k}$ ($\bm{k}'$) is the relative off-shell momentum of the 
two-nucleon system before (after) the interaction $O$.

We differentiate both sides of \Eq{o_me} with respect to $\Lambda$ and
impose the renormalization condition that the matrix element is
invariant under cutoff changes, \IE, $d\braketa{O}/d\Lambda=0$.
This gives the WRG equation for the low-momentum transition operator
(with arguments now explicitly shown),
\begin{eqnarray}
\eqn{rge2}
&&
{\partial O_\Lambda(\bm{k}',\bm{k};p',p) \over\partial\Lambda}
\nonumber\\ 
&& = 
{M} \int {d\Omega_{\hat{\bm{\Lambda}}}\over (2\pi)^3}
\left( {O_\Lambda(\bm{k}',\bm{\Lambda};p',p)
V_\Lambda(\bm{\Lambda},\bm{k};p)\over
1-p^2/\Lambda^2}
\right.
+ \left.{V_\Lambda(\bm{k}',\bm{\Lambda};p')
O_\Lambda(\bm{\Lambda},\bm{k};p',p)\over
1-p'^2/\Lambda^2}
\right) \ ,
\end{eqnarray}
with $\hat{\bm{\Lambda}}\equiv\bm{\Lambda}/\Lambda$.
The low-momentum $NN$-potential, $V_\Lambda(\bm{k}',\bm{k};p)$,
 evolves according to the WRG equation for 
the $NN$ potential~\cite{rg1,birse}. 
The low-momentum operator acquires a dependence on both the initial and final 
on-shell momenta, $p$ and $p'$, as indicated by \Eq{rge2}.

The WRG equation is solved in order to derive $O_\Lambda$ from
$O=O_{\bar{\Lambda}}$ with a cutoff $\bar{\Lambda} (> \Lambda)$.
The solution (in integral form) of the WRG equation 
is (for later convenience now given with partial wave decomposition),
\begin{eqnarray}
\eqn{eff_current}
 O_\Lambda^{(\beta,\alpha)} 
&=& \eta \left( O_{\bar{\Lambda}}^{(\beta,\alpha)} 
+ O_{\bar{\Lambda}}^{(\beta,\alpha)} 
 {1\over E- \lambda H_{\bar{\Lambda}}^{(\alpha)}} 
\lambda V_{\bar{\Lambda}}^{(\alpha)}
+ V_{\bar{\Lambda}}^{(\beta)} \lambda 
{1\over E'-H_{\bar{\Lambda}}^{(\beta)}\lambda} 
 O_{\bar{\Lambda}}^{(\beta,\alpha)}
\right.\nonumber\\
&+& \left. V_{\bar{\Lambda}}^{(\beta)} 
\lambda {1\over E'-H_{\bar{\Lambda}}^{(\beta)}\lambda} 
 O_{\bar{\Lambda}}^{(\beta,\alpha)} 
 {1\over E-\lambda H_{\bar{\Lambda}}^{(\alpha)}}
\lambda V_{\bar{\Lambda}}^{(\alpha)} \right) \eta \ ,
\end{eqnarray}
where $H_{\bar{\Lambda}}^{(\alpha)}$ and $V_{\bar{\Lambda}}^{(\alpha)}$
are the full Hamiltonian and the $NN$-interaction for partial wave $\alpha$, 
defined in the model space with cutoff $\bar{\Lambda}$.
The operator $O_\Lambda^{(\beta,\alpha)}$ generates a transition from a
partial wave $\alpha$ to $\beta$.
The projection operators $\eta$ and $\lambda$ are defined by
\begin{eqnarray}
 \eqn{eta}
\eta &=& \int {\bar{k}^2d\bar{k}\over 2\pi^2} 
\left| \bar{k}\, \right\rangle \left\langle \bar{k}\,
  \right| \,\, ,  \quad \quad \quad
\bar{k}  \leq \Lambda \,\, , \\
\eqn{lambda}
\lambda &=& \int {\bar{k}^2d\bar{k}\over 2\pi^2} 
\left| \bar{k} \,\right\rangle  \left\langle
\bar{k} \, \right| \,\,  ,
\quad \quad \quad
\Lambda < \bar{k} \leq \bar{\Lambda} \,\, ,
\end{eqnarray}
where $\ket{\bar{k}}$ represents the radial part of the free two-nucleon
states with the relative momentum $\bar{k}$.
Equation (\ref{eq:eff_current}) is the same as for the effective operator in 
the Bloch-Horowitz formalism~\cite{bh-np58}.
Similarities between the projection formalisms
(\EG, Bloch-Horowitz and Lee-Suzuki formalisms)
and RG techniques have been explored previously
in other contexts, see, \EG, Refs.~\cite{Bogner,Luu}.

\section{Chiral $s$-wave pion production operator}
\label{sec_op}

Here we present the $\chi$PT-based $s$-wave
pion production operator for the $NN\to d\pi$ reaction near threshold.
We will start our RG analysis with this operator,
and examine the consistency between the RG
running of this chiral operator and the counting rule.
%
%
We use the operator from Ref.~\cite{lensky}, which was 
derived using the modified power counting rule proposed in 
Ref.~\cite{mod_count}.
In this counting the large relative momentum ($p=\sqrt{Mm_\pi}$) of the 
incoming nucleons is counted as an additional large energy-momentum scale of 
the problem, leading to an expansion in $\chi\sim\sqrt{\frac{m_\pi}{M}}$ rather
than in the $\frac{m_\pi}{M}$ of the original Weinberg 
counting~\cite{Weinberg}.
The leading-order (LO) operators are given by the (nucleon recoil) one-body 
operator and rescattering via the Weinberg-Tomozawa (WT) interaction.
At the next-to-leading order (NLO), several loop diagrams start to contribute, 
the sum of which does not vanish in the chiral limit when sandwiched 
between wave functions~\cite{gardestig}, leading to a divergent matrix element.
A solution to this dilemma was proposed in Ref.~\cite{lensky}, where it was 
shown that the WT term and its recoil correction, taken together with a pion 
exchange extracted from the initial or final state wave functions, contribute 
an irreducible NLO diagram that exactly cancels the offending divergence of 
the pion loop diagrams.\footnote{
In an irreducible diagram, the nucleons of any two-nucleon cut are off shell by 
$\sim m_\pi$, \IE, an irreducible diagram cannot be split into 
smaller diagrams with all external nucleons on-shell.
}
We are left with WT rescattering, with its energy dependence replaced 
by the on-shell value, and no CSV term.
The result is that, up to NLO, we consider the (nucleon recoil) one-body
plus a modified WT rescattering term (WT$'$), the latter a factor $4/3$ 
stronger than the original WT term~\cite{lensky}.
These operators are given (in momentum space) by
\begin{eqnarray}
\eqn{sLO}
  O_{\rm WT'} & = & \frac{g_A\omega_q}{4f_\pi^3}
  \varepsilon^{abc}\tau_1^b\tau_2^c
  \left(\frac{\bm\sigma_1\cdot(\bm{k}'_1-\bm{k}_1)}
       {{m_\pi'}^2+(\bm{k}'_1-\bm{k}_1)^2}
       - \frac{\bm\sigma_2\cdot(\bm{k}'_2-\bm{k}_2)}
	    {{m_\pi'}^2+(\bm{k}'_2-\bm{k}_2)^2}\right), \\
\eqn{sLO2}
  O_{\rm 1B} & = & 
  \frac{-ig_A\omega_q(2\pi)^3}{4Mf_\pi}
  \left[\tau_1^a\delta^{(3)}\left(\bm{k}'_2-\bm{k}_2\right)
    \bm\sigma_1\cdot\left(\bm{k}_1+\bm{k}'_1)\right)\right. 
    \nonumber \\ & & 
    \left.+\tau_2^a\delta^{(3)}\left(\bm{k}'_1-\bm{k}_1\right)
    \bm\sigma_2\cdot\left(\bm{k}_2+\bm{k}'_2)\right)\right] \ ,
\end{eqnarray}
where ${m_\pi'}^2=\frac{3}{4}m_\pi^2$ and 
$\omega_q$ is the energy of the emitted pion.
The momentum for $i$-th nucleon in the initial (final) state is denoted by
$\bm{k}_i$ ($\bm{k}'_i$).
The axial-vector coupling constant and the pion decay constant are
denoted by $g_A$ and $f_\pi$, respectively.
We have employed the so-called fixed-kinematics approximation, in which 
the energy transfer is equally shared between the incoming nucleons and
fixed to 
the threshold value $m_\pi/2$, where $m_\pi$ is the pion mass.
Other choices are possible~\cite{pi_energy}, but we will take this simple 
prescription here, and relegate an investigation regarding this issue to 
future work.

When we let the one-body operator run according to the WRG equation,
we obtain a low-momentum operator with a kink
structure. (See Fig.~\ref{fig_run}.)
The origin of the kink
is high momentum components of the bare one-body operator
that are integrated out.
We explain here this point more using the WRG equation [\Eq{rge2}].
When the cutoff is reduced by $\delta\Lambda$,
the running of the operator due to the momentum shell of the
one-body operator integrated out is given up to the order of 
$\delta\Lambda$ as
\begin{eqnarray}
\eqn{rge_one_body}
{\delta O_\Lambda(\bm{k}',\bm{k};p',p)}
= 
{M} \int {d\Omega_{\hat{\bm{\Lambda}}}\over (2\pi)^3}
\left( {O_{{\rm 1B}\;\Lambda}(\bm{k}',\bm{\Lambda})
V_\Lambda(\bm{\Lambda},\bm{k};p)\over
1-p^2/\Lambda^2}
\right.
+ \left.{V_\Lambda(\bm{k}',\bm{\Lambda};p')
O_{{\rm 1B}\;\Lambda}(\bm{\Lambda},\bm{k})\over
1-p'^2/\Lambda^2}
\right)\delta\Lambda \ ,
\end{eqnarray}
where the relative nucleon momenta 
before and after the insertion of $O_{\rm 1B}$ are
$\bm{k} = (\bm{k}_1-\bm{k}_2)/2$ and
$\bm{k}' = (\bm{k}'_1-\bm{k}'_2)/2$, respectively, and they
are related to the pion momentum ($\bm{q}$)
through $\bm{k}'=\bm{k}-\frac{\bm{q}}{2}$.
The first term in r.h.s. of \Eq{rge_one_body}
is non-vanishing only when 
$\bm{k}'+\frac{\bm{q}}{2}=\bm{\Lambda}$ because of the
$\delta$-function in \Eq{sLO2}.
This means that the shift of the operator
[$\delta O_\Lambda(\bm{k}',\bm{k};p',p)$]
is generated only for 
$|\bm{k}'|\ge \Lambda - |\frac{\bm{q}}{2}|$.
Similarly, the second term in r.h.s. of \Eq{rge_one_body}
induces the running of the operator only for
$|\bm{k}|\ge \Lambda - |\frac{\bm{q}}{2}|$;
the kink structure is created in this way.
On the other hand, if $O_{\rm 1B}$ in 
\Eq{rge_one_body} is replaced by 
$O_{\rm WT'}$ [\Eq{sLO}], then 
the shift of the operator
[$\delta O_\Lambda(\bm{k}',\bm{k};p',p)$]
is induced for all values of $\bm{k}$ and $\bm{k}'$ because 
$O_{\rm WT'}$ does not contain the $\delta$-function.
Thus the kink structure does not mean the strong dependence of the
operator on the cutoff. 
Rather, it originates from the $\delta$-function in the one-body
operator, or in other words, from the way we have chosen to define the
transition operators and wave functions. 
We could have chosen a set of transition operators without the one-body 
operator by extracting the one-pion-exchange (or some other) mechanism,
which is nearest the one-body operator,
from the wave function and connecting it to the one-body operator;
a two-body operator is formed in this way. 
The RG running of this operator does not generate the kink structure.
However, we use the one-body operator here because it
is often used in practical calculations and
the kink part will contribute only marginally to the RG running
for the reaction near threshold ($|\bm{q}|\sim 0$) which is of our
interest here. (At $|\bm{q}|=0$, the kink does not appear.)
When fitting the counter terms to the RG low-momentum operator,
we simply omit the kink structure which
obviously cannot be simulated by counter terms.
As we said, the kink part contributes only marginally to the RG running,
and thus we do not consider this omission to be influential on a conclusion
of this work.

\section{Renormalization group running, chiral symmetry and counter
 terms}
\label{sec_counter}

We study the RG running, guided by the WRG equation, of the $\pi$
production operator presented in the previous section.
In $\chi$PT, 
the running coupling
constants of the chiral counter terms should capture the WRG running of
the operator.
%
The leading order counter terms (with one spatial derivative) from the
heavy-baryon $\chi$PT Lagrangian, relevant to this work, are given 
as~\cite{ando_pppi0}
\begin{eqnarray}
\eqn{nnlo}
{\cal L}_{ct} &=&
\frac{i}{2 M f_\pi}
\frac{g_A}{M f_\pi^2}
\left\{ 
\hat{D}_{1a} N^\dagger(\bm{\tau}\cdot\dot{\bm{\pi}} 
\vec{\sigma}\cdot \stackrel{\rightarrow}{\nabla} - 
\vec{\sigma}\cdot \stackrel{\leftarrow}{\nabla}
\bm{\tau}\cdot\dot{\bm{\pi}} 
)NN^\dagger N
\right.  \nonumber \\ &&
+ \hat{D}_{1b}N^\dagger \bm{\tau}\cdot\dot{\bm{\pi}} \vec{\sigma} 
N\cdot N^\dagger 
(\stackrel{\rightarrow}{\nabla}-\stackrel{\leftarrow}{\nabla})N
+\hat{D}_{1c}N^\dagger 
(\bm{\tau}\cdot\dot{\bm{\pi}}
\stackrel{\rightarrow}{\nabla}-\stackrel{\leftarrow}{\nabla}\bm{\tau}\cdot
\dot{\bm{\pi}})N\cdot
N^\dagger \vec{\sigma} N
\nnb \\ &&
+\hat{D}_{1d}N^\dagger \bm{\tau}\cdot\dot{\bm{\pi}} NN^\dagger 
\vec{\sigma}\cdot (\stackrel{\rightarrow}{\nabla} - 
\stackrel{\leftarrow}{\nabla}
)N
+i \hat{D}_{1e} \epsilon_{abc} N^\dagger 
(\bm{\tau}\cdot\dot{\bm{\pi}} 
\stackrel{\rightarrow}{\nabla}_a
+\stackrel{\leftarrow}{\nabla}_a \bm{\tau}\cdot\dot{\bm{\pi}})
\sigma_b 
NN^\dagger \sigma_c N
\nnb \\ &&
+i \hat{D}_{1f} \epsilon_{abc} N^\dagger \bm{\tau}\cdot\dot{\bm{\pi}} 
\sigma_b NN^\dagger \sigma_c 
(\stackrel{\rightarrow}{\nabla}_a + \stackrel{\leftarrow}{\nabla}_a)N
\nnb \\ &&
+\hat{D}_{1g} N^\dagger[\bm{\tau}\cdot\dot{\bm{\pi}} 
\vec{\sigma}\cdot \stackrel{\rightarrow}{\nabla} + 
\vec{\sigma}\cdot \stackrel{\leftarrow}{\nabla}
\bm{\tau}\cdot\dot{\bm{\pi}},\tau_i]NN^\dagger \tau_i N
+ \hat{D}_{1h}N^\dagger [\bm{\tau}\cdot\dot{\bm{\pi}} \vec{\sigma},\tau_i] 
N\cdot N^\dagger 
(\stackrel{\rightarrow}{\nabla}+\stackrel{\leftarrow}{\nabla})\tau_i N
\nnb \\ &&
\left.
+i \hat{D}_{1i} \epsilon_{abc} N^\dagger 
[\bm{\tau}\cdot\dot{\bm{\pi}}
\stackrel{\rightarrow}{\nabla}_a
-\stackrel{\leftarrow}{\nabla}_a \bm{\tau}\cdot\dot{\bm{\pi}},
\tau_i]\sigma_b
NN^\dagger \sigma_c\tau_i N
\right\} \ ,
\end{eqnarray}
where $\hat{D}_{1a}$--$\hat{D}_{1i}$ are dimensionless coupling
constants.
These counter terms are next-to-next-to-leading order (N$^2$LO) terms in
the chiral $s$-wave pion production operator.

We explicitly calculate the WRG running of the 
coefficients of the leading chiral counter terms
($\hat{D}_{1a}$--$\hat{D}_{1i}$ in \Eq{nnlo})
for an infinitesimal reduction of the cutoff;
the LO chiral nuclear force is used together.
The details of the calculation and the result are given in the Appendix.
This calculation also illustrates that
no CSV terms (such as Eq.~(25) in Ref.~\cite{gardestig}
\footnote{
Discussion in the first paragraph of Sec.~\ref{sec_op} gives a reason
why the authors of Ref.~\cite{gardestig} obtained the CSV term.
})
 are generated in the WRG running.
Here, we generalize the result and
exclude a possibility that the WRG running generates a CSV
term, provided that the starting operator does not include a CSV term.
In \Eq{eff_current},
the second, third and fourth terms, which we will refer to as the
generated terms, are to be captured by the counter terms.
This part is linear in the starting operator.
From \Eq{sLO}, we see that the starting operator is proportional to the
pion energy and so are the generated terms.
Thus, the counter terms include the time derivative of the pion field.
Meanwhile,
the starting operator and the full Hamiltonian have symmetries
such as hermiticity, rotational invariance and parity,
and the RG running maintains these symmetries.
To be consistent with the symmetries,
each of the generated terms has to include
a pseudoscalar factor formed by
the spatial derivatives and the nucleon spin operators.
In fact, 
the heavy-baryon $\chi$PT Lagrangian
always includes such counter terms which are 
the most general terms consisting of
the time derivative of the pion field and four nucleon fields.
Thus no CSV terms are generated in the WRG running.

Using the result in the appendix,
we can (roughly) infer the size of the running in case the cutoff is
reduced by a certain finite amount.
Consider the running of the counter terms driven by 
$O_{\rm WT'}$ [\Eq{sLO}] and the one-pion-exchange potential 
($V_{\rm OPEP}$) [\Eq{vpi}].
For an infinitesimal reduction of the cutoff, the running of the counter
terms is given by \Eq{run1}.
Integrating over $\Lambda$ from $\bar{\Lambda}$ down to $\Lambda$ 
($\bar{\Lambda}>\Lambda$) gives
\begin{eqnarray}
\eqn{run1-2}
\hat{D}^{\rm (1-loop)}_{1a} &=& 
{4\; g_A^2 M^3 
\over 3\;(4\pi f_\pi)^2m_\pi}
\left\{\arctan \left(\bar{\Lambda}\over m_\pi\right)
-\arctan \left(\Lambda\over m_\pi\right)
\right\}
 \ .
\end{eqnarray}
In case $\bar{\Lambda}>\Lambda \gg m_\pi$, we have
\begin{eqnarray}
\eqn{run1-4}
\hat{D}^{\rm (1-loop)}_{1a} 
&\simeq &
{4\; g_A^2 M^3 (\bar{\Lambda}-\Lambda)
\over 3\;(4\pi f_\pi)^2 \bar{\Lambda}\Lambda}
 \ .
\end{eqnarray}
The relative strength among $\hat{D}_{1a}$ and other $\hat{D}_{1x}$ 
($x=b,...,i$) is the same as \Eq{run1}.
This is the running of the counter terms generated by the one-loop 
formed by $O_{\rm WT'}$ and $V_{\rm OPEP}$, in which the internal
relative momentum of the two nucleons runs from $\bar{\Lambda}$ to
$\Lambda$.
The running of the counter terms also captures contributions from
the sum of the ladder diagrams in which $O_{\rm WT'}$ is dressed by
multiple iteration of $V_{\rm OPEP}$.
Such a contribution roughly gives the running coupling constant as
\begin{eqnarray}
\eqn{run1-3}
\hat{D}^{\rm (multi-loops)}_{1a} &\sim& 
\hat{D}^{\rm (1-loop)}_{1a}
\left(1+
{M(\bar{\Lambda}-\Lambda)\over (4\pi f_\pi)^2}
+\left({M(\bar{\Lambda}-\Lambda)\over (4\pi f_\pi)^2}\right)^2
+\cdots\right) \nonumber \\
&=& \hat{D}^{\rm (1-loop)}_{1a}
\left( { 1 \over 
1- 
{M(\bar{\Lambda}-\Lambda)\over (4\pi f_\pi)^2}
}\right) \ ,
\end{eqnarray}
where $\bar{\Lambda}-\Lambda < 4\pi f_\pi$ is assumed
from the first to second line.
Although not explicitly shown, 
each term on the R.H.S. of the first line, in fact, 
is accompanied by a dimensionless constant factor
which is ${\cal O}(1)$ and is different for different terms.
Even though \Eq{run1-3} is 
a rather crude estimation of $\hat{D}^{\rm (multi-loops)}_{1a}$,
it still implies a condition for 
the WRG running of the chiral
counter terms to be of natural size:
$\bar{\Lambda}-\Lambda \ll  4\pi f_\pi$.
(We supposed that
the starting chiral counter terms are of natural size.)
If this condition is not satisfied, the running of the counter terms
may not necessarily be finite.~\footnote{
Our argument here has a similarity to that used for identifying
the chiral symmetry breaking scale ($\Lambda_\chi$) with
$\Lambda_\chi \sim 4\pi f_\pi$\cite{georgi}.
Both of the arguments are based on the size of the RG running.
However, we do not think that there is a direct connection between the
two arguments.
In Ref.~\cite{georgi}, a cutoff variation of a reasonable amount leads
to a renormalization of couplings, and the size of the
renormalization ($\sim 1/(4\pi f_\pi)^2$) is related to the chiral symmetry
breaking scale; the RG running 
is due to one-loop diagrams of the $\pi\pi$ interaction (perturbative).
In our case, we found
the size of the cutoff variation ($\sim 4\pi f_\pi$) which may take the
couplings away from the natural size;
the RG running here is due to the sum of the ladder diagrams
of the $NN$ interaction (non-perturbative).
}
Furthermore, we find that $\bar{\Lambda}-\Lambda \ll 4\pi f_\pi$
is a condition to justify the naive expectation that
the WRG running has the dependence on the cutoff as $\propto 1/\Lambda$.
For a perturbative calculation,
this expectation is true,
irrespective of the size of $\bar{\Lambda}-\Lambda$,
as shown in \Eq{run1-4}. 
In a non-perturbative calculation such as $NN$ interaction, however, the
scaling of the coupling is more complicated as seen in \Eq{run1-3}.
Now we estimate the size of the generated coupling constant which
should be of natural size for the counting rule to work.
Taking a physically reasonable range ($\bar{\Lambda}$ = 800~MeV and
$\Lambda$ = 500~MeV), which is fairly consistent with the condition
obtained above ($\bar{\Lambda}-\Lambda \ll 4\pi f_\pi$), 
we obtain with \Eqs{run1-4}{run1-3}
\begin{eqnarray}
\hat{D}^{\rm (1-loop)}_{1a} &\sim& 0.9,\qquad
\hat{D}^{\rm (multi-loops)}_{1a} \sim 1.2 \ ,
\end{eqnarray}
which are the running of natural size.
The above analysis is for the leading order chiral counter terms.
A similar analysis can also be applied to higher order counter terms to
show that the running of the couplings is of natural size.
Thus this qualitative analysis
supports the consistency between the RG running and the
counting rule in case the cutoff is varied within the physically
reasonable range.

In this paragraph, we discuss an implication from
the above analysis for the nuclear force.
Consider the one-pion-exchange potential ($V_{\rm OPEP}$).
The WRG running of $V_{\rm OPEP}$ generates a series of contact
interactions with even numbers of derivatives, \IE,
\begin{eqnarray}
V_{\rm CT} (\bm{k}',\bm{k})= {1\over 4f_\pi^2} \left(\hat{C}_0 +
{\hat{C}_2\over \Lambda^2} (k^2 + k^{\prime\; 2}) + \cdots
\right) \ ,
\end{eqnarray}
where $\hat{C}_i$ ($i=0,2,\cdots$) are dimensionless couplings.
These couplings have a RG running similar to \Eq{run1-3}.
Therefore,
if we reduce the cutoff within the reasonable range (\EG,
$\bar{\Lambda}$ = 800~MeV and $\Lambda$ = 500~MeV),
the resultant renormalization of the couplings is ${\cal O} (1)$, which 
implies consistency between the WRG running and Weinberg's counting.
However, the couplings could be divergent when the cutoff is
varied by $\bar{\Lambda}-\Lambda > 4\pi f_\pi$.
In fact, divergences of the RG running of contact $NN$ interactions
for some partial waves
are indeed observed in Ref.~\cite{nogga} (\EG, see Fig.~2 of the paper)
in which the cutoff is varied over a very wide range from 2 fm$^{-1}$
to 20
 fm$^{-1}$;
\Eq{run1-3} is consistent with the divergent behavior.
Obviously, the counting rule does not work for the divergent contact
interactions.
We vary the cutoff within the physically reasonable range where 
the counting rule is not broken by the divergence,
and study
the consistency between the RG running and the counting rule.

Although the above analysis based on the (semi-)analytic result
suggested that
the RG running of the chiral counter terms is of
natural size, it was a rather crude estimation.
Thus it is desirable to confirm this result by a
numerical calculation as follows.
Starting with the chiral NLO $\pi$-production operator [\Eq{sLO}] with 
$\Lambda = 800$~MeV,
we reduce the cutoff down to $\Lambda = 500$~MeV, with the use
of the integral form of the WRG equation [\Eq{eff_current}].
\footnote{
In this work, we consider a pion production near threshold where the
on-shell relative momentum is about 360~MeV. Therefore, it is impossible
for an operator with a cutoff less than 360~MeV to treat the pion
production.
Also, we should consider explicitly degrees of freedom whose details
matter to a problem under consideration. 
Therefore, we consider 500~MeV to be an appropriate lowest value for the
cutoff. 
}
Then we examine whether
the generated terms as a result of the WRG running are captured by the
chiral counter terms in a way consistent with the counting rule.
This time, we use 
the WRG equation [\Eq{eff_current}] for which 
the partial wave decomposition has been done.
Because we are interested in the $NN\to d\pi$ reaction 
near threshold, we only need to consider the
$^3P_1\to{}^3S_1$-$^3D_1$ transition in the $NN$ system.
Since we use \Eq{eff_current},
the resultant RG effective operator does not have spin and isospin
structure any more, and only the radial dependence is at hand.
In fact, some of the chiral counter terms with different
spin and isospin structures give the same radial dependence 
in our case ($^3P_1\to{}^3S_1$-$^3D_1$ transition near threshold).
Thus we know that the generated part includes various counter terms
[\Eq{nnlo}], but we cannot separate them in practice
because they give the same radial dependence.
What we can do is to use a representative of such counter terms
to parameterize the generated part.
Fortunately, this situation still does not ruin our goal of examining the
consistency between the RG running and the counting rule.
We use the following chiral counter terms as representatives for
parameterizing the generated part:
\begin{eqnarray}
 {\cal L} & = & 
 \frac{\hat{D_{1a}}}{2 M f_{\pi}}\frac{g_A}{M f_{\pi}^2}
 (iN^{\dagger}\vecbox{\tau}\cdot\dot{\vecbox{\pi}}
 \vec{\sigma}\cdot\stackrel{\rightarrow}{\nabla}N {+} H.c.)
 N^{\dagger}N \nonumber\\
&+&  \frac{\hat{D_{3a}}}{2 M f_{\pi}\Lambda^2}\frac{g_A}{M f_{\pi}^2}
 (iN^{\dagger}\vecbox{\tau}\cdot\dot{\vecbox{\pi}}
        \vec{\sigma}\cdot\stackrel{\rightarrow}{\nabla}N {+} H.c.)
(N^{\dagger}\stackrel{\rightarrow}{\nabla}^{2}N  {+} H.c.)\nonumber\\
&+&  \frac{\hat{D_{3b}}}{2 M f_{\pi}\Lambda^2}\frac{g_A}{M f_{\pi}^2}
	(iN^{\dagger}\vecbox{\tau}\cdot\dot{\vecbox{\pi}}
        \vec{\sigma}\cdot\stackrel{\rightarrow}{\nabla}N {+} H.c.)
	(\stackrel{\rightarrow}{\nabla}N^{\dagger}\cdot
	\stackrel{\rightarrow}{\nabla}N) \nonumber\\
&+&  \frac{\hat{D_{3c}}}{2 M f_{\pi}\Lambda}\frac{g_A}{M f_{\pi}^2}
	(iN^{\dagger}\vecbox{\tau}\cdot\dot{\vecbox{\pi}}
        \vec{\sigma}\cdot\stackrel{\rightarrow}{\nabla}N {+} H.c.)
	(N^{\dagger}\biggl( i \stackrel{\rightarrow}{\partial t}+
	    {\stackrel{\rightarrow}{\nabla}^{2}\over 2M}
	    \biggr) N  {+} H.c.)\nonumber\\
&+&  \frac{\hat{D_{5a}}}{2 M f_{\pi}\Lambda^4}\frac{g_A}{M f_{\pi}^2}
 (iN^{\dagger}\vecbox{\tau}\cdot\dot{\vecbox{\pi}}
        \vec{\sigma}\cdot\stackrel{\rightarrow}{\nabla}N {+} H.c.)
	(N^{\dagger} (\;\stackrel{\rightarrow}{\nabla}^{2})^2
N + H.c.) \nonumber\\
&+&  \frac{\hat{D_{5b}}}{2 M f_{\pi}\Lambda^4}\frac{g_A}{M f_{\pi}^2}
	(iN^{\dagger}\vecbox{\tau}\cdot\dot{\vecbox{\pi}}
        \vec{\sigma}\cdot\stackrel{\rightarrow}{\nabla}N {+} H.c.)
	N^{\dagger}(\stackrel{\leftarrow}{\nabla}\cdot
	\stackrel{\rightarrow}{\nabla})	(\stackrel{\rightarrow}{\nabla}^{2}+
	\stackrel{\leftarrow}{\nabla}^{2})N  \nonumber\\
&+&  \frac{\hat{D_{5c}}}{2 M f_{\pi}\Lambda^4}\frac{g_A}{M f_{\pi}^2}
	(iN^{\dagger}\vecbox{\tau}\cdot\dot{\vecbox{\pi}}
        \vec{\sigma}\cdot\stackrel{\rightarrow}{\nabla}N {+} H.c.)
	N^{\dagger}(\stackrel{\leftarrow}{\nabla}\cdot
	\stackrel{\rightarrow}{\nabla})^2N \nonumber\\
&+&  \frac{\hat{D_{5d}}}{2 M f_{\pi}\Lambda^2}\frac{g_A}{M f_{\pi}^2}
	(iN^{\dagger}\vecbox{\tau}\cdot\dot{\vecbox{\pi}}
        \vec{\sigma}\cdot\stackrel{\rightarrow}{\nabla}N {+} H.c.)
	(N^{\dagger}\biggl( i \stackrel{\rightarrow}{\partial t}+
	    {\stackrel{\rightarrow}{\nabla}^{2}\over 2M}
	    \biggr)^2 N  {+} H.c.)\nonumber\\
&+&  \frac{\hat{D_{5e}}}{2 M f_{\pi}\Lambda^3}\frac{g_A}{M f_{\pi}^2}
 (iN^{\dagger}\vecbox{\tau}\cdot\dot{\vecbox{\pi}}
        \vec{\sigma}\cdot\stackrel{\rightarrow}{\nabla}N {+} H.c.)
	(N^{\dagger} (\;\stackrel{\rightarrow}{\nabla}^{2}+
	\stackrel{\leftarrow}{\nabla}^{2})
\biggl(i\stackrel{\rightarrow}{\partial t}
+{\stackrel{\rightarrow}{\nabla}^{2}\over 2M}
\biggr) N + H.c.) \nonumber\\
&+&  \frac{\hat{D_{5f}}}{2 M f_{\pi}\Lambda^3}\frac{g_A}{M f_{\pi}^2}
	(iN^{\dagger}\vecbox{\tau}\cdot\dot{\vecbox{\pi}}
        \vec{\sigma}\cdot\stackrel{\rightarrow}{\nabla}N {+} H.c.)
	(N^{\dagger}(\stackrel{\leftarrow}{\nabla}\cdot
	\stackrel{\rightarrow}{\nabla})	
	\biggl(i\stackrel{\rightarrow}{\partial t}+
	      {\stackrel{\rightarrow}{\nabla}^{2}\over 2M}
	      \biggr)N + H.c.) \ ,
 \eqn{ct}
\end{eqnarray}
where the $\hat{D}_x (x=1a,\cdots,5f)$ are dimensionless coupling constants. 
This is a power expansion with respect to the RG scale $\Lambda$, \IE,
in powers of $p/\Lambda$.
We presented higher order chiral counter terms with three or five
derivatives.
Although there are also many spin-isospin structures possible 
for the counter terms with three or five 
derivatives, we presented only the representatives which we indeed use
in parameterizing the generated part.
We do not have a criterion to select the set of operators in \Eq{ct}.
Even if we included some more operators, we would not have a better
parameterization. 
The coefficients for those counter terms are
$\hat{D}_{3a}$--$\hat{D}_{3c}$ and
$\hat{D}_{5a}$--$\hat{D}_{5f}$, where the
index 3 and 5 indicate how many nucleon derivatives they have.
Note that contact operators with an even number of spatial derivatives 
will not survive sandwiching between wave functions for the
$^3P_1\to{}^3S_1$-$^3D_1$ transition.

The counter terms with the LECs $\hat{D}_{3c}$, $\hat{D}_{5d}$,
$\hat{D}_{5e}$, and $\hat{D}_{5f}$ are the so-called redundant terms.
The redundant terms 
are generated in the WRG running,
as explicitly shown in the Appendix.
Although those terms may be eliminated by a field redefinition,
the field redefinition is accompanied by a contribution from the 
measure (the Jacobian factor) which is difficult to calculate~\cite{harada}.
Thus, we explicitly consider the redundant terms, which means that we
explicitly consider the on-shell energy dependence of the low-momentum 
operator.

\section{Numerical Results}
\label{sec_result}
We start with the chiral NLO $s$-wave pion production operator for the
$NN\to d\pi$ reaction, presented in Sec~\ref{sec_op}.
The starting operator is defined in the model space with 
$\Lambda = 800$~MeV (sharp cutoff).
Using the integral form of the WRG equation in \Eq{eff_current}, 
we calculate the RG low-momentum operator for $\Lambda=500$~MeV.\footnote{
We may use either \Eq{rge2} after the partial wave decomposition,
or \Eq{eff_current} to calculate an effective
RG operator; the result is the same. We use \Eq{eff_current} for an
easier calculation.
}
Unless otherwise stated, we use the low-momentum $NN$ potential obtained
from the CD-Bonn $NN$-potential~\cite{cdbonn} 
to generate the initial proton-neutron $^3P_1$ scattering and
final deuteron wave functions.
Some reasons for using the phenomenological nuclear force rather than a
$\chi$PT-based nuclear force have been given in the introduction.
For the purposes of this paper we can ignore any charge-dependent effects and
assume that our calculation (of $np\to d\pi^0$) applies equally well to 
(Coulomb-corrected) $pp\to d\pi^+$.
Thus we will consistently write $NN\to d\pi$.
We also employ near threshold kinematics, \IE, $\eta\sim0.1$,
where $\eta=q/m_\pi$ is the emitted pion momentum divided by the pion mass.
The chiral counter terms that we use to simulate the generated part of the
RG low-momentum operator have been presented in \Eq{ct}.

\subsection{The $^3P_1\to{}^3S_1$ transition}
\begin{figure}[t]
\includegraphics[width=6in]{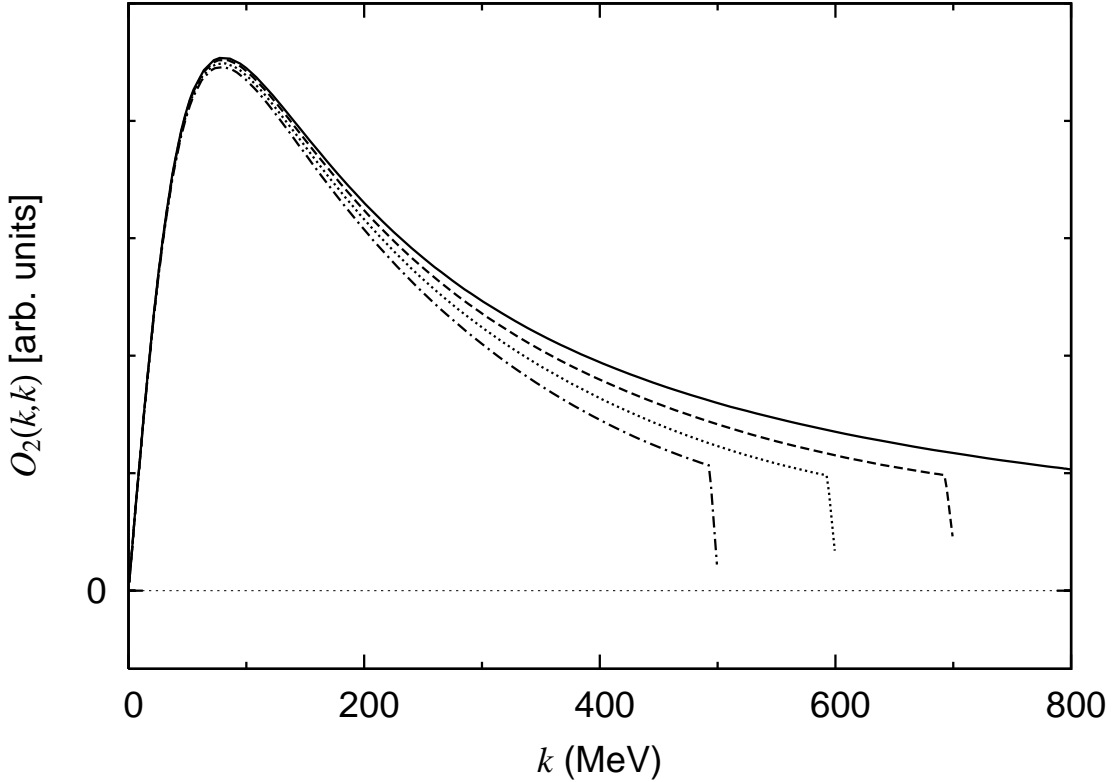}
\caption{\label{fig_run}
Running of the pion production operator for the $^3P_1\to{}^3S_1$ transition 
in $NN\to d\pi$.
The diagonal momentum space matrix elements are shown.
The starting chiral NLO operator ($\Lambda=800$~MeV)
is shown by the solid line.
After the RG running, we obtain the low-momentum operator with 
$\Lambda=700$~MeV (dashed line), 
$\Lambda=600$~MeV (dotted line), 
and $\Lambda=500$~MeV (dash-dotted line); $\eta = 0.1$.
The abrupt drop in the curves close to the cutoff is a reflection of our
treatment of the one-body operator (the kink structure).}
\end{figure}

The running of the radial part of the $^3P_1\to {}^3S_1$ transition operator 
(diagonal matrix elements) is shown in Fig.~\ref{fig_run}.
The solid line is the starting NLO chiral operator (before RG running).
After the RG running, we obtain the RG low-momentum operator shown by the 
dashed line ($\Lambda$~=~700~MeV),
dotted line ($\Lambda$~=~600~MeV) and
dash-dotted line ($\Lambda$~=~500~MeV).
Only the result for the diagonal components is given here---a similar trend is 
found for the off-diagonal components.
As stated earlier, we expect to observe a kink structure, which is
similar to the ``jump-up'' structure in Ref.~\cite{NA}, in the
low-momentum operators.
Here it actually appears up as an abrupt drop close to the cutoff.

Now we parameterize this low-momentum operator using the NLO operator plus
the counter terms in \Eq{ct}. 
We omit the kink part when fitting the counter terms.
Since the low-momentum operator has the on-shell energy dependence,
we also need to include the redundant terms in our fit.
The low-momentum operator is calculated for a range in
$\eta$ between 0.02 and 0.1 with steps of 0.02.
We fit the LECs to these low-momentum operators using the least squares method 
and with both diagonal and off-diagonal matrix elements as input.
The $\eta$-dependence is then parameterized by the redundant terms.
All momenta smaller than $\Lambda$ are used with equal weight 
in the fitting.\footnote{
In principle the smaller momentum should have larger weight in the fitting,
since in an effective theory the low-momentum behavior of the
operator is more accurately described. 
However, as we will see shortly, we can easily achieve a good fit for the 
full range $<\Lambda$.
Therefore we do not think it necessary to put heavier
weight on the smaller momenta.
}
The LECs obtained in this way are presented in Table~\ref{tab_Ds}.
We notice that $\hat{D}_{1a}$ changes significantly when adding the 
$\hat{D}_{3x}$ terms.
This is a consequence of introducing the redundant term $\hat{D}_{3c}$.
A similar change is observed also in $\hat{D}_{3a}$, $\hat{D}_{3b}$, and
$\hat{D}_{3c}$ as the $\hat{D}_{5x}$ terms are added.
The LECs in Table~\ref{tab_Ds} are used to obtain the results in the
following figures and tables.

In Fig.~\ref{fig_diag}, we show the simulation of the low-momentum
operator using the NLO operator plus the counter terms.
The dash-dotted line is the starting NLO chiral operator and the solid
line is the low-momentum operator.
\begin{figure}[t]
\includegraphics[width=6in]{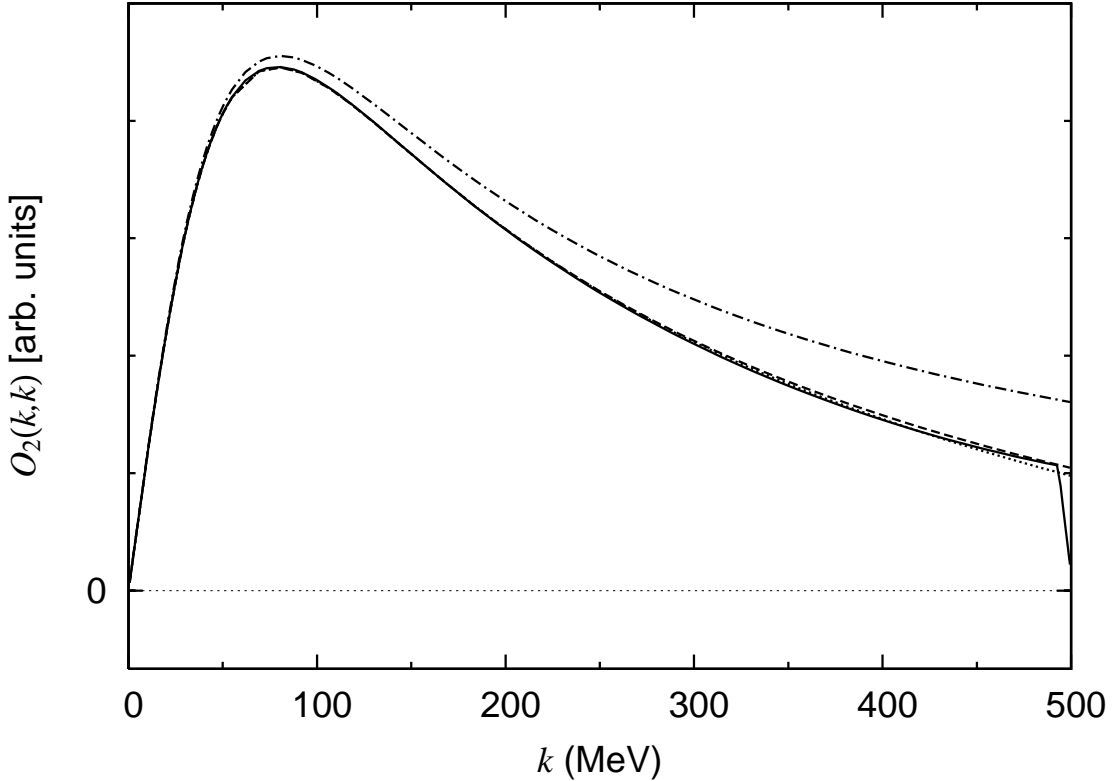}
\caption{\label{fig_diag}
Simulation of the low-momentum pion production operator
($\Lambda$ = 500 MeV, $\eta$ = 0.1) 
for the $^3P_1\to{}^3S_1$ transition in $NN\to d\pi$.
The diagonal matrix elements are shown.
The low-momentum operator (solid line) is simulated by the
original NLO operator (one-body and rescattering terms, dash-dotted line) 
plus a contact term with one nucleon derivative (dashed line).
The dotted line has additional contact terms
with three derivatives.}
\end{figure}
The dashed line contains the NLO operator plus the counter term with
coefficient $\hat{D}_{1a}$.
The dotted line contains the additional terms $\hat{D}_{3a}$, $\hat{D}_{3b}$,
and $\hat{D}_{3c}$ of \Eq{ct}. 
The generated part (the difference between the solid and dash-dotted lines) 
is thus accurately captured by the higher order counter terms.
In addition, the size of the dimensionless couplings is natural
as can be gleaned from 
Table~\ref{tab_Ds}, where we also include five derivatives counter
terms.\footnote{
The LECs $\hat{D}_{5c}$ and $\hat{D}_{5f}$ are relatively large,
although not disturbingly so.
However, because of the operator structure, 
the $\hat{D}_{5c}$-term appears with a factor $1/3$ in the $^3P_1\to {}^3S_1$
transition matrix element. 
The $\hat{D}_{5f}$-term has a factor $\Lambda/M$, compared to the 
$\hat{D}_{x}$-terms ($x = 5a, 5b, 5c$), which makes its contribution smaller by
a factor of $\sim 2$.
Therefore, the seemingly large LECs are ascribable to the result of the 
coefficients which accompany the LECs.
}
\begin{table}[t]
\renewcommand{\arraystretch}{1.2}
\tabcolsep=3.5mm
\caption{\label{tab_Ds}
The coupling constants $\hat{D}_x$ for $\Lambda$ = 500~MeV.
The first column is the transition which the counter
 terms induce. In the second column, ``1'' means that we use the counter
 term with one nucleon derivative in simulating the low-momentum operator.
For ``3'' (``5''), we additionally use the counter terms with three
 (three and five) derivatives.}
\begin{tabular}{cccccccc}\hline
&\# of $\nabla$ & $\hat{D}_{1a}$ & $\hat{D}_{3a}$ & $\hat{D}_{3b}$ & 
  $\hat{D}_{3c}$&$\hat{D}_{5a}$
\\ \hline
\multirow{3}{18mm}[0mm]{$^3P_1\to {}^3S_1$}
& 1&  0.77& -& -&  -&-\\
& 3&  0.39& $-$0.64&  0.70&  1.31&-\\
& 5&  0.42& $-$0.24&  1.98&  0.44&  0.52\\\hline
\multirow{2}{20mm}[0mm]{$^3P_1\to {}^3D_1$}
&3 & -&-&$-$1.37&-&-\\		   
&5 & -&-&$-$1.21&-&-\\\hline\hline
&\# of $\nabla$ & $\hat{D}_{5b}$ & $\hat{D}_{5c}$ & $\hat{D}_{5d}$ & 
  $\hat{D}_{5e}$&$\hat{D}_{5f}$
\\ \hline
\multirow{1}{18mm}[0mm]{$^3P_1\to {}^3S_1$}
& 5& 5.63&7.63& 2.39&$-$2.08&$-$8.19\\\hline
\multirow{1}{20mm}[0mm]{$^3P_1\to {}^3D_1$}
&5&0.93& $-$1.62&-&-&$-$1.58\\\hline
\end{tabular}
\end{table}
The natural sizes of the LECs show that the chiral expansion of the
contact operator indeed converges very well, confirming the visual information 
in Fig.~\ref{fig_diag}.

\begin{figure}[t]
\includegraphics[width=6in]{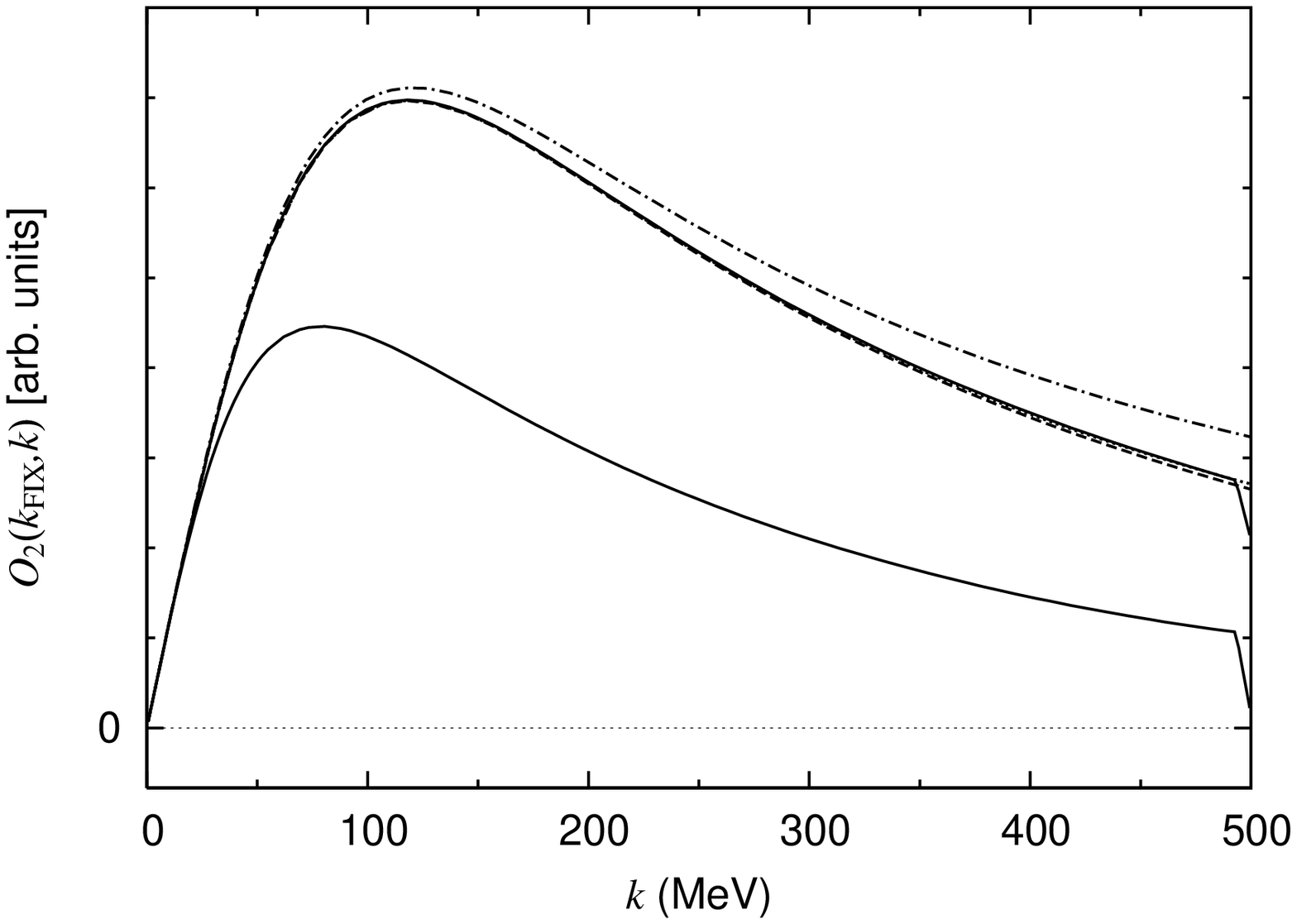}
\caption{\label{fig_off_diag}
Simulation of the low-momentum pion production operator
($\Lambda$ = 500 MeV, $\eta$ = 0.1) for the $^3P_1\to{}^3S_1$ transition
 in $NN\to d\pi$.
The off-diagonal momentum space matrix elements are shown for
$k_{\rm FIX}$ = 10~MeV.
The diagonal components are also shown by the lower solid line.
The other features are the same as in Fig.~\ref{fig_diag}.}
\end{figure}

Because of the large momentum mismatch between the initial and final on-shell 
momenta in $NN\to d\pi$, the off-diagonal components are actually more 
important than the diagonal ones.
The results for the off-diagonal matrix elements are shown in 
Fig.~\ref{fig_off_diag}.
The low-momentum RG operator is again well parameterized using the higher 
order counter terms.
A similar trend is seen also for the other off-diagonal components.

\subsection{The $^3P_1\to {}^3D_1$ transition}
\begin{figure}[t]
\includegraphics[width=6in]{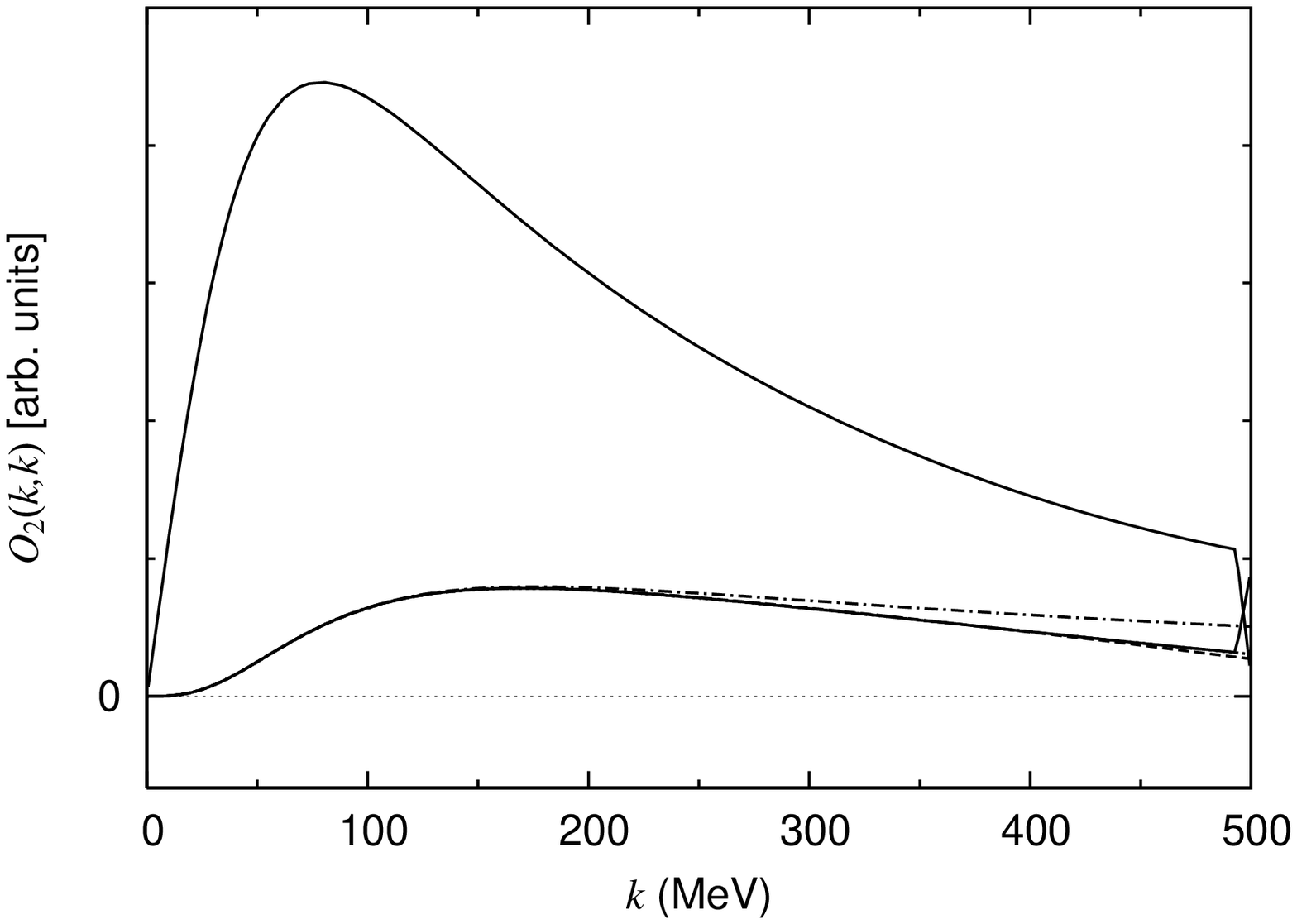}
\caption{\label{fig_diag_3d1}
Simulation of the low-momentum pion production operator 
($\Lambda$ = 500 MeV, $\eta$ = 0.1) 
for the $^3P_1\to{}^3D_1$ transition in $NN\to d\pi$.
The diagonal momentum space matrix elements are shown.
The low-momentum operator (lower solid line) is simulated by the original
NLO operator (one-body and rescattering terms, dash-dotted line) plus a
contact term with the $\hat{D}_{3b}$ coefficient (dashed line).
The dotted line, which almost falls on the solid line, additionally
includes counter terms with five derivatives.
The operator for the $^3P_1\to{}^3S_1$ transition
is shown for comparison (upper solid line).}
\end{figure}

We also examine the running of the operator for the $^3P_1\to {}^3D_1$
transition. 
The result is shown in Fig.~\ref{fig_diag_3d1},
where the diagonal matrix elements of the operator are given.
A counter term needs to have at least three derivatives
for this transition, so the counter term 
($\hat{D}_{1a}$) in \Eq{ct} does not contribute. 
We simulate the low-momentum operator using the counter term 
($\hat{D}_{3b}$)---the $\hat{D}_{3a}$ and $\hat{D}_{3c}$
 terms cannot contribute.
We choose the LECs in the same way as described in the previous subsection.
This fit can be done independently of the fit of the $^3P_1\to{}^3S_1$ 
transition operator since the other counter terms not shown in \Eq{ct} now
contribute in a different linear combination.
Already the one-parameter-fit (dashed line) is fairly accurate.
When we include the counter terms with five derivatives, we obtain an almost 
perfect simulation as shown by the dotted line in Fig.~\ref{fig_diag_3d1}.
Also these $\hat{D}_x$ are natural (see Table~\ref{tab_Ds}).
The same level of accuracy is also observed in the simulation of the
off-diagonal components.

\subsection{Matrix Elements}

We present in Table~\ref{tab_ct_amp} the matrix elements of the
parameterized low-momentum operator to show the accuracy of the
parametrization using the counter terms.
\begin{table}[t]
\caption{\label{tab_ct_amp}
Matrix elements of the pion production operator 
including the one-body, rescattering and several counter terms 
($\eta = 0.1$).
In the first row,
``0'' includes no counter terms, i.e., it refers to the original chiral 
NLO operator.
``1'' includes the counter term with one nucleon derivative, while
``3'' (``5'') includes additional counter terms with three (three and five) 
derivatives. 
All matrix elements are divided by the matrix element of the RG
low-momentum operator with the kink omitted, to which the counter terms
 are fitted.}
\renewcommand{\arraystretch}{1.2}
\tabcolsep=3.5mm
 \begin{tabular}{ccccc}\hline
$\Lambda$ (MeV)& 0& 1 & 3 & 5 \\\hline
 700& 1.02& 1.01& 1.00& 1.00\\	     
 600& 1.05& 1.01& 1.00& 1.01\\	     
 500& 1.11& 1.01& 1.00& 1.00\\\hline
 \end{tabular}
\end{table}
It is clear that the counter terms accurately absorb the generated part of 
the low-momentum operator---already with the first counter term the
expansion deviates from the low-momentum operator with only 1\%.
With further higher order counter terms included, 
the matrix elements of the low-momentum operator and its parametrization
become practically indistinguishable.
The convergence of the counter term expansion is thus very good.

In Table~\ref{tab_amp}, the values for the matrix element of each component of 
the operator are given to show the significance of the contribution from
the generated terms.
\begin{table}[t]
\caption{\label{tab_amp}
Relative contributions to the matrix element from each component of the pion
production operator for the original operator ($\Lambda=800$~MeV) and a few 
other cutoffs; $\eta=0.1$.
The symbols, ``1B'', ``WT$^\prime$'' and ``Generated term'' denote 
contributions from the one-body, modified WT term, and generated terms, 
respectively.
The sum of all contributions is normalized to unity.}
\renewcommand{\arraystretch}{1.2}
\tabcolsep=3.5mm
\begin{tabular}{cccc}\hline
$\Lambda$ (MeV)& $\braketa{\rm 1B}$&$\braketa{\rm WT'}$&
  $\braketa{\rm Generated\ term}$\\ \hline
 800& 0.07& 0.93& 0\\
 700& 0.07& 0.95&-0.02\\      
 600& 0.05& 1.01&-0.05\\      
 500& 0.01& 1.09&-0.11\\\hline
\end{tabular}
\end{table}
At $\Lambda = 500$~MeV, the generated part contributes by as much as 11\%.
This significant contribution for the relatively small cutoff 
($\Lambda=500$~MeV) can be expected, because the cutoff is fairly close to 
the on-shell momentum of the initial $NN$ state 
($p\sim\sqrt{Mm_\pi}\approx360$~MeV).
Thus, $NN$ states which considerably contribute to the reaction have 
been integrated out and their strength shifted to contact terms.

\subsection{Wave-function dependence}
\begin{figure}[t]
\begin{center}
\includegraphics[width=6in]{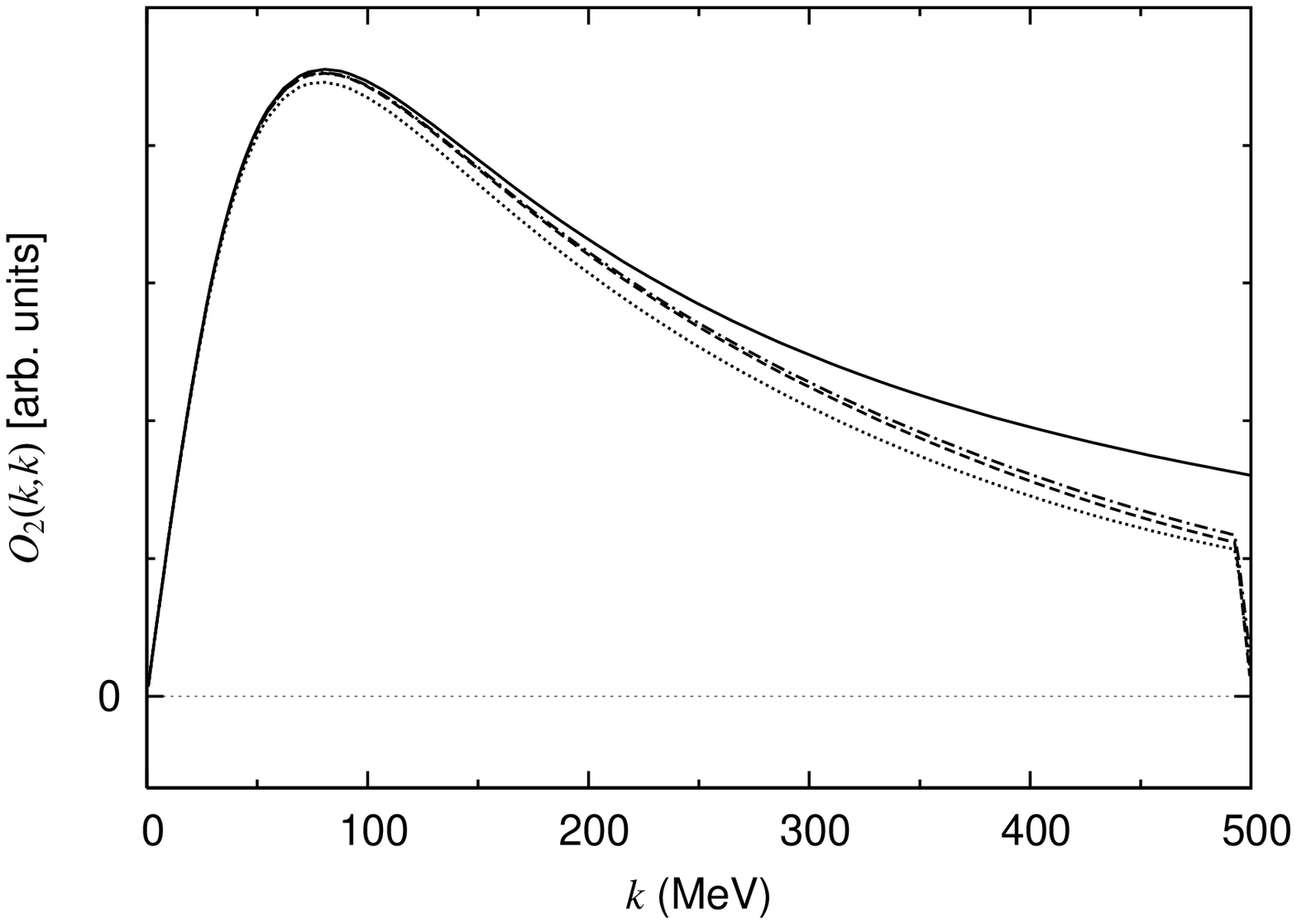}
\caption{\label{fig_av18}
Dependence of the running of the pion production operator
($^3P_1\to{}^3S_1$) on the choice of
nuclear force for diagonal momentum space matrix elements.
The original chiral NLO operator (solid line) evolves to the dashed (AV18), 
dash-dotted line (Nijmegen I), or the dotted (CD-Bonn) line. 
The cutoff value for the low-momentum operators is $\Lambda$ = 500~MeV;
$\eta$ = 0.1.}
\end{center}
\end{figure}

In this subsection we investigate the sensitivity of our result to
variations in the $NN$ potential.
Using the AV18~\cite{av18} and the Nijmegen I~\cite{nij} potentials,
instead of 
the CD-Bonn potential, we again study the running of the chiral NLO 
operator.
The result is shown in Fig.~\ref{fig_av18} together with the earlier result 
for the CD-Bonn potential for comparison.
Obviously, the NLO operator evolves to different low-momentum
operators, depending on the choice of nuclear potentials.
This result is understandable since the matrix element of the chiral NLO
operator ($\Lambda = 800$~MeV) rather depends on the choice of the
nuclear force, as demonstrated in Table~\ref{tab:NN-dep}.
\begin{table}[t]
\renewcommand{\arraystretch}{1.2}
\tabcolsep=3.5mm
\caption{\label{tab:NN-dep}
The reduced $s$-wave cross section,
$\alpha=\sigma/\eta|_{\eta=0}$, for various $NN$ potentials.}
\begin{tabular}[t]{cccc}\hline
 & CD-Bonn & AV18 & Nijmegen I \\
$\alpha$ ($\mu$b)& 237& 205& 224\\\hline
\end{tabular}
\end{table}
However, the low-momentum operators can be accurately parameterized by
the chiral NLO operator plus several counter terms with the natural
strength, just as we have seen in
the case of the CD-Bonn potential.
The accuracy of the parametrization is at a similar level
regardless of potentials.

The discrepancy between the matrix elements calculated by different potentials
can be regarded as a higher order effect.
This is supported by the size of the differences in the cross section
(Table~\ref{tab:NN-dep}), and is also confirmed by the operators---the spread
in the curves of Fig.~\ref{fig_run} stemming from RG running is similar to the 
spread in Fig.~\ref{fig_av18} between different wave functions.

In the numerical analysis presented above,
the counter terms allowed by the chiral symmetry accurately simulate the 
generated part of the RG low-momentum operator, with the coefficients of
the natural strength.
We have examined the accuracy of the fitting both graphically and
quantitatively (Table~\ref{tab_ct_amp}).
Thus, our result allows us to conclude that
the RG running is consistent with the counting rule
for the cutoff changed over the physically reasonable range.
This result is also consistent with our (rough) semi-analytic estimation
presented in Sec.~\ref{sec_counter}.

%
%

\section{Summary}\label{sec_summary}

In nuclear $\chi$PT with the cutoff regularization,
the cutoff has a physical meaning and its choice is not arbitrary,
however, it still may be varied within a reasonable range.
We examined whether the running of the counter
terms, which is a result of the cutoff variation, is consistent with
the counting rule.
This consistency is a condition for $\chi$PT to be useful.
We proposed to use Wilsonian RG (WRG) equation for this investigation.
With the use of WRG equation, 
it is guaranteed that
no chiral-symmetry-violating (CSV) terms are generated, provided that we
start with an operator
consistent with the chiral symmetry.
Besides, the RG running of the operator, which exactly reflects the high
momentum states integrated out, is at hand, and thus
we can study the convergence of the chiral expansion of the RG running.

As a demonstration, we applied the WRG equation to the
$s$-wave $\pi$-production operator. 
We started with the chiral NLO operator~\cite{lensky} with the sharp cutoff,
$\Lambda=800$~MeV.
We showed that no CSV terms are generated in the WRG running by
analytically calculating the WRG running of the pion production operator
for the infinitesimal reduction of the cutoff;
the chiral LO nuclear force is used together.
In this calculation, we used differential form of the WRG equation without
the partial wave decomposition.
We made the rough estimate
based on the result of this analytic calculation, 
and found a range of the cutoff variation for which
the WRG running is of natural size.
We also found that
the WRG running may be divergent 
when the cutoff is changed more than 1~GeV;
the WRG running does not necessarily have a simple $\propto 1/\Lambda$
behavior when the cutoff is changed by $\gtap$1~GeV.
Therefore, for the variation of the cutoff over
the physically reasonable range,
the RG running is consistent with the counting rule.
This argument based on the analytic calculation
is also applicable to the nuclear force and other transition operators.
Thus we made a remark that the WRG running of the nuclear force is also
consistent with Weinberg's counting rule, as long as we are concerned with a
variation of the cutoff within the reasonable range.

In order to confirm the rather rough estimate given above,
we numerically calculated the running of the chiral counter terms for
the variation of the cutoff over the physically reasonable range.
The CD-Bonn potential is used in this calculation.
The couplings of the counter terms were fitted to 
the RG running of the operator given by the WRG equation.
We found that the parametrization in terms of the counter terms
is indeed accurate.
Thus, the matrix elements for the low-momentum operator can be reproduced 
within 1\% already by including the NLO operator plus just the first 
(N$^2$LO) counter term, with
further improvements when higher order counter terms are included. 
All the couplings of the
counter terms have natural strength, and thus the running can be
safely considered to be of higher order than the NLO.
We also used different phenomenological $NN$ potentials, and
found that the result is essentially the same as above.
As mentioned in the introduction,
this result implies that the phenomenological $NN$ potentials used do
not contain a significant CSV component, and thus
would serve as a support of the validity of the hybrid approach.
The remaining difference in the matrix element of the NLO operator
among these $NN$ potentials can most likely be captured
by the N$^2$LO counter terms, some of which would be fitted directly to data.
Thus our result indicates that,
if the cutoff is varied within the physically reasonable range,
the WRG running and the counting rule are consistent, at least for this
reaction and to the order we are working.

\begin{acknowledgments}
The author acknowledges Anders G{\aa}rdestig for many useful discussions.
He also thanks to Koji Harada and Shung-ichi Ando for informative discussions.
This work was supported by the Natural Science and Engineering Research Council
of Canada and Universidade de S\~ao Paulo.
\end{acknowledgments}

\appendix*

\section{Wilsonian renormalization group running of chiral counter terms}

Here we show analytically that 
the running,
based on the Wilsonian renormalization group (WRG) equation,
of a pion production operator is captured by the
renormalization of the chiral counter terms,
provided that we start with a pion production 
operator consistent with chiral symmetry; 
the operator vanishes at
threshold and the chiral limit.
We use the differential form of the WRG equation [\Eq{rge2}] 
to study what operators are generated as a result of the RG running.
We use here the WRG equation without partial wave decomposition in order to
easily identify the operator structure of the resultant operators.
We examine the running of the chiral NLO $s$-wave
$\pi$ production operator given in \Eq{sLO}.
For the $NN$ interaction, we use the chiral LO $NN$ interaction, given as:
\begin{eqnarray}
 \eqn{vpi}
 V_\Lambda(\bm{k}',\bm{k};p) 
&=& - \bm{\tau}_1\cdot\bm{\tau}_2 \,\frac{g_A^2}{4 f_\pi^2}\,
\frac{ \bm{\sigma}_1\cdot (\bm{k}'_1-\bm{k}_1)
\,\bm{\sigma}_2\cdot (\bm{k}_2-\bm{k}'_2)
}
{(\bm{k}'_1-\bm{k}_1)^2+m_\pi^2} 
+ C_S (\Lambda) + C_T (\Lambda)\;\bm{\sigma}_1\cdot\bm{\sigma}_2, \nonumber \\
\end{eqnarray}
with $\bm{k}'_1-\bm{k}_1=\bm{k}_2-\bm{k}'_2 = \bm{k}'-\bm{k}$;
$\bm{k}_i(\bm{k}'_i)$ is the momentum of the $i$-th nucleon before (after) the
interaction.
The first term is the one-pion-exchange potential ($V_{\rm OPEP}$) and the
second and third terms are contact interactions with different
dependences on the nuclear spin.
The couplings of the contact interactions are functions of $\Lambda$.
We start with the set of the $NN$ potential and the $\pi$ production
operator specified above.
The operators are defined in the model space with the cutoff $\Lambda$.
We reduce $\Lambda$ by an infinitesimal amount $|\delta\Lambda|$, and see
the running of the $\pi$ production operator, $\delta O$.
In the following, we replace $O_\Lambda$ in \Eq{rge2} with either
$O_{\rm WT'}$ or $O_{\rm 1B}$, and $V_\Lambda$ with either the 
$V_{\rm OPEP}$ or one of the contact interactions (we use the $C_S$-term only),
and calculate $\delta O$ for each combination of $O$ and $V$.
We will see that $\delta O$ is captured by the renormalization of the
chiral counter terms, except for the kink part from the running of
$O_{\rm 1B}$.
We show this only for the leading counter terms which have one spatial
derivative. 
We do not explicitly calculate further higher order counter terms from
$\delta O$.
Also, our calculation of $\delta O$ does not exhaust every possible
combination of $O$ and $V$.
Nevertheless, our presentation is enough for our purpose of
demonstrating how the RG running is captured by the chiral
counter terms.
As stated in the text, no chiral-symmetry-violating terms are 
generated by the RG running anyway.
We perform the following calculation for the threshold kinematics.
Above the threshold, 
the starting operators in \Eq{sLO} have the dependence on the
center-of-mass momentum even when we set the initial state in the center
of mass system,
and thus the resultant generated term also has the dependence on it.
This dependence leads to an
additional renormalization of the chiral
counter terms.

\subsection{$O = O_{\rm WT'}$ and $V = V_{\rm OPEP}$}
\vspace{-10mm}
\begin{eqnarray}
\eqn{delta_o1}
&&\delta O^{(1)}_\Lambda(\bm{k}',\bm{k};p',p) =
-|\delta\Lambda|{M} 
\left(-{g_A^2\over 4f_\pi^2}\right)
\left({g_A m_\pi \over 4f_\pi^3}\right)
\int {d\Omega_{\hat{\bm{\Lambda}}}\over (2\pi)^3}\nonumber\\
&\times&\left( 
\frac{\bm\sigma_1\cdot(\bm{k}'_1- \bm{\Lambda})
\bm{\sigma}_1\cdot (\bm{\Lambda}-\bm{k}_1)\,
\bm{\sigma}_2\cdot (\bm{\Lambda}+\bm{k}_2)
}
{[{m_\pi'}^2+(\bm{k}'_1 - \bm{\Lambda})^2]
[m_\pi^2 + (\bm{\Lambda}-\bm{k}_1)^2]
(1-p^2/\Lambda^2)
}
\epsilon_{abc}\tau_1^b\tau_2^c
(\bm{\tau}_1\cdot\bm{\tau}_2)
\right. \nonumber\\
&+& \left.\frac{ \bm{\sigma}_1\cdot (\bm{k}'_1-\bm{\Lambda})\,
\bm{\sigma}_2\cdot (-\bm{k}'_2-\bm{\Lambda})
\bm\sigma_1\cdot(\bm{\Lambda}-\bm{k}_1)
}
{[m_\pi^2 + (\bm{k}'_1-\bm{\Lambda})^2]
[{m_\pi'}^2+(\bm{\Lambda}-\bm{k}_1)^2]
(1-p'^2/\Lambda^2)}
(\bm{\tau}_1\cdot\bm{\tau}_2)
\epsilon_{abc}\tau_1^b\tau_2^c
\right) 
+ (1\leftrightarrow 2),
\end{eqnarray}
where the factor $(-|\delta\Lambda|)$ means the infinitesimal 
{\it reduction} of the cutoff.
$\delta O^{(1)}_\Lambda$ is captured by a series of the
counter terms.
In order to see this more clearly,
we expand the factors in the denominator as follows:
\begin{eqnarray}
\eqn{expand1}
{1\over m_\pi^2 + (\bm{k}-\bm{\Lambda})^2}
&=& 
{1\over m_\pi^2 + \Lambda^2 + k^2 -2\bm{\Lambda}\cdot\bm{k}}\nonumber\\
&=& 
{1\over m_\pi^2 + \Lambda^2}
-{k^2 -2\bm{\Lambda}\cdot\bm{k}
\over (m_\pi^2 + \Lambda^2)^2}
 + \cdots \ ,
\end{eqnarray}
\begin{eqnarray}
\eqn{expand2}
{1\over 1-p^2/\Lambda^2}
&=& 1 + p^2/\Lambda^2 - \cdots\ ,
\end{eqnarray}
and so on.
The expansion in \Eq{expand2} generates the on-shell momentum dependent
terms which are to be captured by the redundant counter terms.
After the expansion of \Eqs{expand1}{expand2}, we perform the angular
integral of $\hat{\bm{\Lambda}}$, eliminating the terms
with odd numbers of $\bm{\Lambda}$.
Here, we just keep terms contributing to the renormalization of the
leading counter terms shown in \Eq{nnlo}.
That is, 
we retain the leading terms of the expansion 
in terms of $1/(m_\pi^2 + \Lambda^2)$ in \Eq{delta_o1}, 
and take the first term in \Eq{expand2}.
Note that the second term in the second line of \Eq{expand1} also
gives a contribution.
Thus we have,
\begin{eqnarray}
\eqn{delta_o1_2}
\delta O^{(1)}_\Lambda(\bm{k}',\bm{k};p',p) &=&
-{|\delta\Lambda|{M} \over 6\pi^2 ({m_\pi}^2+\Lambda^2)}
\left(-{g_A^2\over 4f_\pi^2}\right)
\left({g_A m_\pi \over 4f_\pi^3}\right)
\nonumber\\
&\times&\bigg[
\{- 2 \bm\sigma_1\cdot(\bm{k}_1-\bm{k}'_1)
+ \bm\sigma_1\cdot(\bm{k}_2-\bm{k}'_2)
\}
\epsilon_{abc}\tau_1^b\tau_2^c
\nonumber\\
&+& 
\{ 2 \bm\sigma_1\cdot(\bm{k}_1+\bm{k}'_1)
- \bm\sigma_2\cdot(\bm{k}_1+\bm{k}'_1)
\}
(2i) (\tau_1^a - \tau_2^a)
\bigg]
+ (1\leftrightarrow 2).
\end{eqnarray}
The terms generated in this way are captured by the renormalization of
the chiral counter terms in \Eq{nnlo} as follows:
\begin{eqnarray}
\eqn{run1}
\delta\hat{D}_{1a} = 
2\delta\hat{D}_{1b} = 
-2\delta\hat{D}_{1c} = 
-\delta\hat{D}_{1d} = 
2\delta\hat{D}_{1g} = 
-4\delta\hat{D}_{1h} = 
{4\; g_A^2 M^3|\delta\Lambda|\over
3\;(4\pi f_\pi)^2(m_\pi^2+\Lambda^2)}  \ .
\end{eqnarray}

\subsection{$O = O_{\rm WT'}$ and $V = C_{S}$}
\vspace{-10mm}
\begin{eqnarray}
\eqn{delta_o2}
\delta O^{(2)}_\Lambda(\bm{k}',\bm{k};p',p) &=&
-|\delta\Lambda|{M} 
C_S(\Lambda)
\left({g_A m_\pi \over 4f_\pi^3}\right)
\int {d\Omega_{\hat{\bm{\Lambda}}}\over (2\pi)^3}
\left( 
\frac{\bm\sigma_1\cdot(\bm{k}'_1- \bm{\Lambda})}
{[{m_\pi'}^2+(\bm{k}'_1 - \bm{\Lambda})^2]
(1-p^2/\Lambda^2)
}
\right. \nonumber\\
%
&+& \left.
\frac{\bm\sigma_1\cdot(\bm{\Lambda}-\bm{k}_1)}
{[{m_\pi'}^2+(\bm{\Lambda}-\bm{k}_1)^2]
(1-p'^2/\Lambda^2)}
\right) 
\epsilon_{abc}\tau_1^b\tau_2^c 
+ (1\leftrightarrow 2)
\ .
\end{eqnarray}
As done for $\delta O^{(1)}_\Lambda$,
we keep the leading contributions
[$1/(m_\pi'^2+\Lambda^2)$] to obtain
\begin{eqnarray}
\eqn{delta_o2_2}
\delta O^{(2)}_\Lambda(\bm{k}',\bm{k};p',p) &=&
{|\delta\Lambda|{M} C_S(\Lambda)
 \over 6\pi^2 ({m_\pi'}^2+\Lambda^2)}
\left({g_A m_\pi \over 4f_\pi^3}\right)
{\bm\sigma_1\cdot(\bm{k}_1- \bm{k}'_1)}
\epsilon_{abc}\tau_1^b\tau_2^c 
+ (1\leftrightarrow 2) \ .
\end{eqnarray}
This generated term is captured by the renormalization of the chiral
counter terms:
\begin{eqnarray}
\delta\hat{D}_{1g} = 
{C_S(\Lambda) M^3|\delta\Lambda|\over
24\pi^2 (m_\pi^{\prime 2} +\Lambda^2)}  \ .
\end{eqnarray}

\subsection{$O = O_{\rm 1B}$ and $V = V_{\rm OPEP}$}
\vspace{-10mm}
\begin{eqnarray}
\eqn{delta_o3}
&&\delta O^{(3)}_\Lambda(\bm{k}',\bm{k};p',p) =
-|\delta\Lambda|{M} 
\left(-{g_A^2\over 4f_\pi^2}\right)
\left({-i g_A m_\pi (2\pi)^3 
\over 4 M f_\pi}\right)
\nonumber\\
&\times&
\int {d\Omega_{\hat{\bm{\Lambda}}}\over (2\pi)^3}
\left( \delta^{(3)}(\bm{k}'_2+\bm{\Lambda})
\frac{\bm\sigma_1\!\cdot\!(\bm{k}'_1+ \bm{\Lambda})
\bm{\sigma}_1\!\cdot\! (\bm{\Lambda}-\bm{k}_1)\,
\bm{\sigma}_2\!\cdot\! (\bm{\Lambda}+\bm{k}_2)
}
{[m_\pi^2 + (\bm{\Lambda}-\bm{k}_1)^2]
(1-p^2/\Lambda^2)
}
\tau_1^a (\bm{\tau}_1\!\cdot\!\bm{\tau}_2)
\right. \nonumber\\
&+& \delta^{(3)}(-\bm{\Lambda}-\bm{k}_2)
\left.\frac{ \bm{\sigma}_1\!\cdot\! (\bm{k}'_1-\bm{\Lambda})\,
\bm{\sigma}_2\!\cdot\! (-\bm{k}'_2-\bm{\Lambda})
\bm\sigma_1\!\cdot\!(\bm{\Lambda}+\bm{k}_1)
}
{[m_\pi^2 + (\bm{k}'_1-\bm{\Lambda})^2]
(1-p'^2/\Lambda^2)}
(\bm{\tau}_1\!\cdot\!\bm{\tau}_2)
\tau_1^a
\right) 
+ (1\leftrightarrow 2) .
\end{eqnarray}
We perform the angular integral ($\hat{\bm{\Lambda}}$) to obtain
\begin{eqnarray}
\eqn{delta_o3_2}
&&\delta O^{(3)}_\Lambda(\bm{k}',\bm{k};p',p) =
-{|\delta\Lambda|{M} \over \Lambda^2}
\left(-{g_A^2\over 4f_\pi^2}\right)
\left({-i g_A m_\pi 
\over 4 M f_\pi}\right)
\nonumber\\
&\times&
\left( \delta(k'_2-\Lambda)
\frac{\bm\sigma_1\!\cdot\!(\bm{k}'_1-{\bm{k}'_2})
\bm{\sigma}_1\!\cdot\! (-{\bm{k}'_2}-\bm{k}_1)\,
\bm{\sigma}_2\!\cdot\! (-{\bm{k}'_2}+\bm{k}_2)
}
{m_\pi^2 + ({\bm{k}'_2}+\bm{k}_1)^2
}
\tau_1^a (\bm{\tau}_1\!\cdot\!\bm{\tau}_2)
\right. \nonumber\\
&+& \delta(\Lambda-k_2)
\left.\frac{ \bm{\sigma}_1\!\cdot\! (\bm{k}'_1+{\bm{k}_2})\,
\bm{\sigma}_2\!\cdot\! (-\bm{k}'_2+{\bm{k}_2})
\bm\sigma_1\!\cdot\!(-{\bm{k}_2}+\bm{k}_1)
}
{m_\pi^2 + (\bm{k}'_1+{\bm{k}_2})^2
}
(\bm{\tau}_1\!\cdot\!\bm{\tau}_2)
\tau_1^a
\right) 
+ (1\leftrightarrow 2)
\ .
\end{eqnarray}
We obtained the operators with the $\delta$-function. 
This term gives the kink structure found in
the text, and is not captured by the chiral counter terms.
We will see later that the RG running of 
$\delta O^{(3)}_\Lambda$ generates the chiral counter terms.

\subsection{$O = O_{\rm 1B}$ and $V = C_{S}$}
\vspace{-10mm}
\begin{eqnarray}
\eqn{delta_o4}
\delta O^{(4)}_\Lambda(\bm{k}',\bm{k};p',p) &=&
-|\delta\Lambda|{M} C_S(\Lambda)
\left({-i g_A m_\pi (2\pi)^3 
\over 4 M f_\pi}\right)
\int {d\Omega_{\hat{\bm{\Lambda}}}\over (2\pi)^3}
\left( \delta^{(3)}(\bm{k}'_2+\bm{\Lambda})
\frac{\bm\sigma_1\cdot(\bm{k}'_1+ \bm{\Lambda})}
{1-p^2/\Lambda^2}
\right. \nonumber\\
&&+ 
\delta^{(3)}(-\bm{\Lambda}-\bm{k}_2)
\left.
\frac{ \bm\sigma_1\cdot(\bm{\Lambda}+\bm{k}_1)}
{1-p'^2/\Lambda^2}
\right) 
\tau_1^a
+ (1\leftrightarrow 2)
\nonumber \\
&=&
-{|\delta\Lambda|{M} C_S(\Lambda) \over \Lambda^2}
\left({-i g_A m_\pi 
\over 4 M f_\pi}\right)
\left( \delta({k}'_2-{\Lambda})
{\bm\sigma_1\cdot(\bm{k}'_1-{\bm{k}'_2})}
\right. \nonumber\\
&&+ 
\delta({\Lambda}-{k}_2)
\left.
{ \bm\sigma_1\cdot(\bm{k}_1-{\bm{k}}_2)}
\right) 
\tau_1^a
+ (1\leftrightarrow 2) \ .
\end{eqnarray}

\subsection{$O = \delta O^{(3)}_\Lambda$ and $V = C_{S}$}
In \Eqs{delta_o3_2}{delta_o4}, we obtained the operators with the
$\delta$-function.
The use of the WRG equation [\Eq{rge2}] once more gives the counter
terms as follows:
\begin{eqnarray}
\eqn{delta_o5}
&&\delta O^{(5)}_\Lambda(\bm{k}',\bm{k};p',p) =
-{M} C_S(\Lambda)
(-){|\delta\Lambda|{M} \over \Lambda^2}
\left(-{g_A^2\over 4f_\pi^2}\right)
\left({-i g_A m_\pi \over 4 M f_\pi}\right)
\int {d\Omega_{\hat{\bm{\Lambda}}}\over (2\pi)^3} \nonumber\\
&\times&\left( |\delta\Lambda|\delta(k'_2-\Lambda)
\frac{\bm\sigma_1\cdot(\bm{k}'_1-{\bm{k}'_2})
\bm{\sigma}_1\cdot (-{\bm{k}'_2}-\bm{\Lambda})\,
\bm{\sigma}_2\cdot (-{\bm{k}'_2}-\bm{\Lambda})
}
{[m_\pi^2 + ({\bm{k}'_2}+\bm{\Lambda})^2]
(1-p^2/\Lambda^2)
}
\tau_1^a (\bm{\tau}_1\cdot\bm{\tau}_2)
\right. \nonumber\\
&+& 
\left.\frac{ \bm{\sigma}_1\cdot (\bm{k}'_1-{\bm{\Lambda}})\,
\bm{\sigma}_2\cdot (-\bm{k}'_2-{\bm{\Lambda}})
\bm\sigma_1\cdot(2{\bm{\Lambda}})
}
{[m_\pi^2 + (\bm{k}'_1-{\bm{\Lambda}})^2]
(1-p^2/\Lambda^2)
}
(\bm{\tau}_1\cdot\bm{\tau}_2)
\tau_1^a
\right. \nonumber \\
&+& \frac{\bm\sigma_1\cdot(2\bm{\Lambda})
\bm{\sigma}_1\cdot ({\bm{\Lambda}}-\bm{k}_1)\,
\bm{\sigma}_2\cdot ({\bm{\Lambda}}+\bm{k}_2)
}
{[m_\pi^2 + (-{\bm{\Lambda}}+\bm{k}_1)^2]
(1-p'^2/\Lambda^2)
}
\tau_1^a (\bm{\tau}_1\cdot\bm{\tau}_2)
\nonumber\\
&+& |\delta\Lambda|\delta(\Lambda-k_2)
\left.\frac{ \bm{\sigma}_1\!\cdot\! (\bm{\Lambda}+{\bm{k}_2})\,
\bm{\sigma}_2\!\cdot\! (\bm{\Lambda}+{\bm{k}_2})
\bm\sigma_1\!\cdot\!(\bm{k}_1-{\bm{k}_2})
}
{[m_\pi^2 + (\bm{\Lambda}+{\bm{k}_2})^2]
(1-p'^2/\Lambda^2)
}
(\bm{\tau}_1\cdot\bm{\tau}_2)
\tau_1^a
\right) 
+ (1\leftrightarrow 2)
\ .
\end{eqnarray}
We perform the angular integral and keep the leading terms
[$\propto 1/(m_\pi^2+\Lambda^2)$] to obtain
\begin{eqnarray}
\eqn{delta_o5_2}
&&\delta O^{(5)}_\Lambda(\bm{k}',\bm{k};p',p) =
{|\delta\Lambda|{M}^2 C_S(\Lambda)\over 6\pi^2 (m_\pi^2 + \Lambda^2)}
\left(-{g_A^2\over 4f_\pi^2}\right)
\left({-i g_A m_\pi \over 4 M f_\pi}\right)
\nonumber\\
&\times&\Big[ |\delta\Lambda|\delta(k'_2-\Lambda)
\{\bm\sigma_1\cdot(\bm{k}'_1-{\bm{k}'_2})
\bm{\sigma}_1\cdot \bm{\sigma}_2
\}
\tau_1^a (\bm{\tau}_1\cdot\bm{\tau}_2)
\nonumber\\
&+& |\delta\Lambda|\delta(\Lambda-k_2)
\{
\bm{\sigma}_1\cdot \bm{\sigma}_2
\bm\sigma_1\cdot(\bm{k}_1-{\bm{k}_2})
\}
(\bm{\tau}_1\cdot\bm{\tau}_2)
\tau_1^a
\nonumber \\
&-& \{ 2\bm{\sigma}_2\cdot (\bm{k}_1-\bm{k}'_1)\,
+6\bm{\sigma}_2\cdot (\bm{k}_2-\bm{k}'_2)\,
-2i (\bm{\sigma}_1\times\bm{\sigma}_2)\cdot (\bm{k}_1+\bm{k}'_1)
\}
i (\bm{\tau}_1\times\bm{\tau}_2)^a
\nonumber\\
&+& \{ 2\bm{\sigma}_2\cdot (\bm{k}_1+\bm{k}'_1)\,
+6\bm{\sigma}_2\cdot (\bm{k}_2+\bm{k}'_2)\,
-2i (\bm{\sigma}_1\times\bm{\sigma}_2)\cdot (\bm{k}_1-\bm{k}'_1)
\}
\tau_2^a
\Big]
+ (1\leftrightarrow 2)
\ .
\end{eqnarray}
The terms without $\delta$-function contribute to the renormalization of
the counter terms as follows:
\begin{eqnarray}
\delta\hat{D}_{1a} = 
3\delta\hat{D}_{1b} = 
3\delta\hat{D}_{1f} = 
-2\delta\hat{D}_{1g} = 
6\delta\hat{D}_{1h} = 
-6\delta\hat{D}_{1i} = 
{|\delta\Lambda| M^3 C_S (\Lambda) g_A^2 \over
8\pi^2 (m_\pi^2 + \Lambda^2)} \ .
\end{eqnarray}

\bibliographystyle{unsrt}

\end{document}